\documentclass[aps,prb,reprint,superscriptaddress,longbibliography]{revtex4-2}
\usepackage{epsfig}
\usepackage{gensymb}
\usepackage{dcolumn}
\usepackage{physics}
\usepackage{graphicx,graphics}
\usepackage{amsmath}
\usepackage{amssymb}
\usepackage{bm}
\usepackage{color}
\usepackage[utf8]{inputenc}
\usepackage[colorlinks,linkcolor={blue},citecolor={blue},urlcolor={blue}]{hyperref}
\usepackage{mathtools}
\usepackage{booktabs}
\usepackage{float}
\usepackage{mathdots}

\begin{document}
  \title{Theory of Tunneling between Two-Dimensional Electron Layers \\ Driven by Spin Pumping: Adiabatic Regime and Beyond}   

\author{Modi Ke}
\affiliation{Department of Physics and Astronomy, The University of Alabama, Tuscaloosa, AL 35487, USA}
\author{Mahmoud M. Asmar}
\affiliation{Department of Physics, Kennesaw State University, Marietta, Georgia 30060, USA}
\author{Wang-Kong Tse}
\affiliation{Department of Physics and Astronomy, The University of Alabama, Tuscaloosa, AL 35487, USA}

\begin{abstract}

%
Tunneling spectroscopy between parallel two-dimensional (2D) electronic systems provides
a powerful method to probe the underlying electronic properties by measuring tunneling
conductance. In this work, we present a theoretical framework for spin transport in
2D-to-2D tunneling systems, driven by spin pumping. This theory applies to a vertical
heterostructure where two layers of metallic 2D electron systems are separated by an
insulating barrier, with one layer exchange-coupled to a magnetic layer driven at resonance.
Utilizing a non-perturbative Floquet-Keldysh formalism, we derive general expressions for
the tunneling spin and charge currents across a broad range of driving frequencies,
extending beyond the traditional adiabatic pumping regime. At low frequencies, we obtain
analytical results that recover the known behaviors in the adiabatic regime. However, at
higher frequencies, our numerical findings reveal significant deviations in the dependence
of spin and charge currents on both frequency and precession angle. This work offers fresh
insights into the role of magnetization dynamics in tunneling transport, opening up new
avenues for exploring non-adiabatic spin pumping phenomena.
\end{abstract}

\maketitle
\section{Introduction}  

Quantum tunneling of electrons between two layers of two-dimensional electron gas (2DEG), as realized firstly in semiconductor double quantum wells \cite{sctunneling1,sctunneling2,sctunneling3,sctunneling4,sctunneling5} and more recently in van der Waals heterostructures \cite{vtunneling1,vtunneling2,vtunneling3,vtunneling4}, has provided a useful probe for the fundamental electronic properties of these two-dimensional (2D) systems. In these setups, an interlayer bias voltage drives a tunneling charge current. In-plane momentum conservation greatly restricts the available phase space for the 2D-to-2D tunneling, and measurement of the tunneling conductance grants unique access to the quasiparticle spectral function and lifetime. 
%
%

It is of fundamental and practical interest for spintronics \cite{RevSpintronics} to investigate the possibility of the 2D-to-2D tunneling of a pure \textit{spin} current across such a heterostructure. In this work, we propose a magnetically-coupled tunneling heterostructure, which consists of an additional magnetic layer exchange-coupled to one of the 2DEG layers in the standard tunneling setup. The magnetic layer is driven at resonance, providing a precessing magnetization field that couples to the 2DEG's electronic spin degrees of freedom. The induced spin precession in that 2DEG then becomes a source of spin pumping, which drives a spin current across the two 2DEG layers in the heterostructure. 


%
%

Conventionally, spin pumping has played a crucial role in the generation and manipulation of spin currents. 
One common mechanism of spin pumping involves exciting a ferromagnetic material with microwave radiation that keeps the ferromagnet's magnetization in precession. 
The precessing magnetization transfers spin angular momentum into an adjacent non-magnetic material, resulting in a spin current. The efficiency of this angular momentum transfer is influenced by different factors such as the spin mixing conductance at the interface, the Gilbert damping constant, and the spin diffusion length ~\cite{Nick2,efficiency,efficiency2}. In the non-magnetic material, the injected spin current leads to a spin accumulation~\cite{Nick1}, 
providing a direct indication of the spin pumping efficiency. Traditionally, spin pumping has been extensively studied as a means for spin injection in lateral heterostructures~\cite{Nick1,Nick2,Nick3,spininj,revpump}. On the other hand, the possibility of employing spin pumping to drive a vertical spin current across a 2D-to-2D tunneling heterostructure, where in-plane momentum conservation plays a profound role, has been unexplored so far.

It is therefore the purpose of this paper to investigate this curious possibility and to elucidate the behavior of such a tunneling spin current. As we will show, this tunneling process is accompanied by a very small vertical charge current, rendering the spin current not $100\%$ polarized. We calculate the time-averaged values of both currents over a magnetization precession cycle and analyze their dependence with respect to various parameters, including the precession angle, driving frequency, and the interfacial exchange coupling between the 2DEG and the magnetic layer. One crucial aspect of our theory is that it is non-perturbative with respect to the driving frequency. This is facilitated by taking into account the exact time-periodic dynamics of the magnetization field by employing the Floquet-Keldysh formalism. 
While it is known that conventional spin pumping theories formulated within the adiabatic low-frequency regime works adequately for ferromagnetic resonance frequencies in the range of $\sim 1\,-\,10\,\mathrm{GHz}$, the Floquet-Keldysh formulation utilized in our present work goes beyond the adiabatic approximation ~\cite{tserk1,tserk2,revpump} and can extend the conventional theory to higher driving frequencies, making it suitable to describe spin pumping from antiferromagnets as well that typically have a much higher resonance frequency reaching the terahertz region. 

The rest of our paper is organized as follows. In Sec.~\ref{Formulation}, we introduce the model for the tunneling heterostructure driven by a precessing magnetization and derive the corresponding Floquet Hamiltonian and Floquet Green's functions. We then detail the theoretical framework for the tunneling spin and charge currents between the two 2DEG layers in Sec.~\ref{currentsec} using the Floquet-Keldysh  formalism. In Sec.~\ref{numerics} the numerical results of the spin and charge currents and their dependence on various system parameters are discussed, and approximate analytical results are  derived up to leading order in the driving frequency. 
Finally Sec.~\ref{conclusion} concludes our paper.

\section{Model}\label{Formulation}

\begin{figure}
   \begin{center}
            \includegraphics[width=\columnwidth]{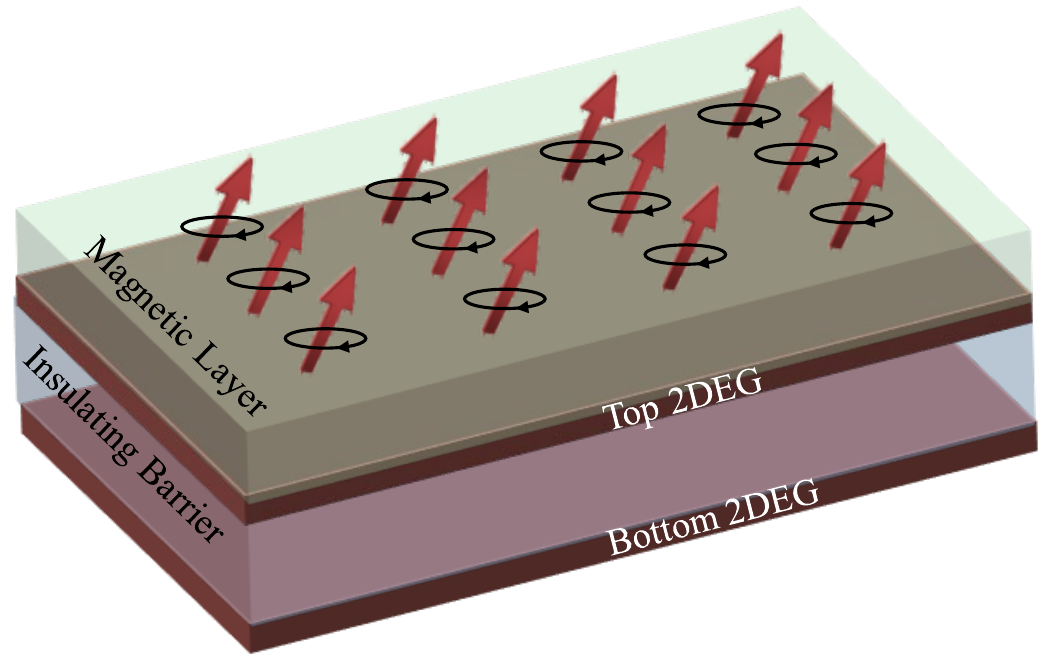}
                \end{center}
                \caption{Setup of the double-layer heterostructure, which consists of two 2DEG layers sandwiching an insulating barrier. The top layer is proximity coupled to a magnetic layer, which is driven at resonance.} \label{setup}

\end{figure}

As shown in Fig.~\ref{setup}, the system setup consists of two layers of 2DEG 
separated by a tunneling barrier, forming a double-layer heterostructure. The $z$-axis is taken as the out-of-plane direction. The top-layer 2DEG is proximity-coupled to an insulating magnetic material, which can be a ferromagnet or A-type antiferromagnet (\textit{e.g.}, $\rm{CrSb}$, $\rm{MnBi_{2n}Te_{3n+1}}$, $\rm{NaCrTe_{2}}$)~\cite{AFM3,AFM4,Mac,AFM5,AFM6}. By suitable tuning of a \textit{d.c.} magnetic field applied along the $z$-direction and simultaneous irradiation of an \textit{a.c.} magnetic field, 
the ferromagnetic or antiferromagnetic material is driven at resonance [ferromagnetic resonance (FMR) or antiferromagnetic resonance (AFMR), respectively], which results in a steady-state precession of its magnetization vector about the $z$-axis. Through interfacial exchange coupling, the top-layer 2DEG's electron spins interact  with this precessing magnetization of the magnetic layer, leading to a time-periodic magnetically driven system.  
%
%

In the following, we develop our theory by taking the 2DEG layers specifically as n-doped 2D  transition metal dichalcogenides (TMDs). As explained below, this will also incorporate the case of 2D electron systems with a parabolic band dispersion as a special limit. 
With the Fermi energy in the TMD's conduction band, the valence band can be projected out using Löwdin’s partitioning leading to an effective Hamiltonian for the conduction band electrons~\cite{TMD}, 
\begin{eqnarray} \label{TMD_Ham_1}
H_{\rm TMD}=\frac{\hbar^2q^2}{2m}+\tau\frac{\lambda }{\Delta} \frac{\hbar^2q^2}{2m}\hat{s}_z,
\end{eqnarray}
where $m$ is the effective mass of the conduction band parabolic dispersion, which is related to the TMD band velocity $v$, band gap $\Delta$ and spin-orbit coupling strength $\lambda$ \cite{TMD2} by $m=(\Delta^2-\lambda^2)/(2\Delta v^2)$. The case of an ordinary 2DEG with a parabolic energy dispersion can be obtained by taking the limit  $\lambda = 0$ and ignoring the valley degrees of freedom in our theory. 
%
%
In the second quantized form, the Hamiltonian of the top-layer electrons interacting with the precessing magnetization is then given by 
\begin{eqnarray}\label{topHam}
H_{\rm T}(t)=\sum_{qs\tau}\varepsilon^0_{qs\tau}\hat{d}^{\dagger}_{qs\tau}\hat{d}_{qs\tau}
+J_{\rm T}\sum_{qss'\tau}\hat{d}^{\dagger}_{qs\tau}[\bar{m}(t)\cdot \bar{\sigma}]_{ss'}\hat{d}_{qs'\tau},\nonumber\\
\end{eqnarray}
where $\hat{d}^{\dagger}_{qs\tau},\hat{d}_{qs\tau}$ are the creation and annihilation operators of the top-layer electrons, and $q,s,\tau$ denote the momentum, spin and valley degrees of freedom respectively. Following from Eq.~\eqref{TMD_Ham_1}, the TMD electronic energy dispersion is given by
\begin{eqnarray}\label{Eqstau}
\varepsilon^0_{qs\tau}=\frac{\hbar^2q^2}{2m}\left(1+\frac{\lambda }{\Delta}\tau s\right).
\end{eqnarray}

The time dependence of the Hamiltonian comes from the exchange coupling with coupling constant $J_{\rm T}$ to the magnetic moment $\bar{m}(t)=(\sin{\alpha}\cos{\Omega t},\sin{\alpha}\sin{\Omega t},\cos{\alpha})$ of the magnetic layer, where the precession angle $\alpha$ is the angle between the magnetization and the $z$-axis. The bottom layer has the same Hamiltonian as Eq.~\eqref{topHam} but without the time-dependent external driving
\begin{eqnarray}\label{botHam}
&&H_{\rm B}=\sum_{qs\tau}\varepsilon^0_{qs\tau}\hat{c}^{\dagger}_{qs\tau}\hat{c}_{qs\tau},
\end{eqnarray}
where $\hat{c}^{\dagger}_{qs\tau}\hat{c}_{qs\tau}$ are the creation and annihilation operators of the bottom-layer electrons. In Eqs.~\eqref{topHam}-\eqref{botHam}, we have ignored a Zeeman coupling term to the external \textit{d.c.} magnetic field   (necessary to drive the magnetization vector into precession), because the Zeeman splitting $\lesssim 10^{-2}\,\mathrm{meV}$ is negligibly small for typical fields $\lesssim 1\,\mathrm{T}$. 

We employ the standard assumptions for 2D-to-2D tunneling assuming that the tunneling amplitude between the two layers of the heterostructure to be  uniform across the entire sample area, and independent of the spin and valley degrees  of freedom~\cite{tunnel,TunnelS}. This uniform tunneling ensures that the in-plane momentum of electrons is preserved during the tunneling process. The tunneling Hamiltonian is thus given by~\cite{tunnel}
\begin{eqnarray}
&&H_{\rm I}=\sum_{qs\tau\tau'}\delta_{\tau,\tau'}(V\hat{c}^\dagger_{qs\tau}\hat{d}_{qs\tau'}+V^*\hat{d}^\dagger_{qs\tau'}\hat{c}_{qs\tau}),
\end{eqnarray}
where $V$ is the tunneling amplitude between the two TMD layers.

\section{Floquet Green's Functions}\label{FGF}

The non-equilibrium retarded $G^R$ and lesser $G^<$ Green's functions are respectively defined by ~\cite{Jahuo,Jauhobook} 
\begin{equation}
G^R_{\alpha,\alpha'}(t,t')=-i\theta(t-t')\langle\{c_\alpha(t),c_{\alpha'}^\dagger(t')\}\rangle,
\end{equation}
\begin{equation}
G^<_{\alpha,\alpha'}(t,t')=i\langle c_{\alpha'}^\dagger(t')c_\alpha(t)\rangle,
\end{equation}
where the bracket $\langle \dots \rangle$ denotes  quantum statistical average. The non-equilibrium Green's functions are not translationally invariant in time, and thus they depend on both time variables. 

The task of taking into account the exact periodic driving dynamics can be facilitated by using the Floquet formalism \cite{Floq2,Floq3,Floq4}. The retarded, advanced and lesser Green's functions can be written in the Floquet representation as  \cite{Floq2},
\begin{eqnarray}
[G(\bm r,\bm r',\bar{\omega})]_{mn}&=&
\\
\frac{1}{T}\int^T_0 d t_{\textrm{av}}\int^{\infty}_{-\infty}&d t_{\textrm{rel}}& e^{i(\bar{\omega}+m\Omega)t-i(\bar{\omega}+n\Omega)t'}G(\bm r,t;\bm r',t'), \nonumber
\end{eqnarray}
where $t_{\textrm{av}}=(t+t')/2$ and $t_{\textrm{rel}}=t'-t$ are the average time and relative time, respectively, and  $\bar\omega \in (-\Omega/2,\Omega/2]$ is the frequency in the reduced zone. 

The system is assumed to be in contact with an external fermionic bath that provides a mechanism for thermalization under time-periodic driving. 
We take the wide-band approximation where the broadening function characterizing the quasiparticle lifetime due to the bath is given by  $\Gamma(\omega)=\Gamma$~\cite{Floq2,wide,wide2}, corresponding to a retarded self-energy $[\Sigma^R(\bar\omega)]_{mn} = -i\Gamma\delta_{m,n}$ 
and a lesser self-energy 
\begin{eqnarray}
[{\Sigma}^{<}(\bar\omega)]_{mn} = 2i\Gamma  f(\bar\omega+m\Omega)\delta_{m,n},
\end{eqnarray}
where $f(\omega)=1/[e^{(\omega-\mu)/k_BT}+1]$ is the Fermi distribution with $\mu$ being the chemical potential. 

After the initial transients have washed out, the system dynamics settles into a non-equilibrium steady state (NESS). In NESS, the lesser Floquet 
Green's function is given by the retarded and advanced Floquet Green's functions as
\begin{eqnarray} \label{lesserG}
[{G}^<(\bar\omega)]_{mn}=\sum_{m'n'}[G^R(\bar\omega)]_{mm'}[\Sigma^<(\bar\omega)]_{m'n'}[G^A(\bar\omega)]_{n'n}.\nonumber\\
\end{eqnarray}
%


The Floquet Hamiltonian is defined by $[\mathcal{H}_{F}]_{mn}=\mathcal{H}_{mn}-n\hbar \Omega \delta_{m,n}$ \cite{Floq1,Floq2}, where the Floquet matrix $\mathcal{H}_{mn}$ is given by
\begin{equation}
\mathcal{H}_{mn}=\frac{1}{T}\int^{T}_{0}dt e^{i(m-n)\Omega t}\mathcal{H}(t).
\end{equation}
%
From Eqs.~\eqref{topHam}-\eqref{Eqstau}, the top-layer Floquet Hamiltonian at valley $\tau$ is thus 
\begin{widetext} 
\begin{eqnarray}
&&H^{\rm T}_{F,\tau}=\begin{bmatrix}
\ddots&\vdots&\vdots&\vdots&\vdots&\vdots&\vdots&\iddots\\
 \dots&\varepsilon_{q\uparrow\tau}+\Omega    
   &  0& 0& 0& 0& 0&\dots\\\dots&0 &  \varepsilon_{q\downarrow\tau}+\Omega
 & J_{\rm T} \sin{\alpha}& 0& 0& 0&\dots\\\dots&0 &  J_{\rm T} \sin{\alpha}
 & \varepsilon_{q\uparrow\tau}& 0& 0& 0&\dots\\\dots&0 & 0
 & 0& \varepsilon_{q\downarrow\tau}&J_{\rm T} \sin{\alpha}& 0&\dots\\\dots&0 & 0
 & 0& J_{\rm T} \sin{\alpha}&\varepsilon_{q\uparrow\tau}-\Omega& 0&\dots\\\dots&0 & 0
 & 0& 0&0& \varepsilon_{q\downarrow\tau}-\Omega\\ \iddots&\vdots&\vdots&\vdots&\vdots&\vdots&\vdots&\ddots\end{bmatrix}, \label{widematrix}
\end{eqnarray}
\end{widetext}
where $\varepsilon_{qs\tau}=\varepsilon_{qs\tau}^0+s J_{\rm T} \cos{\alpha}$ and $s\in \{\uparrow,\downarrow\}$ labels the spin degree of freedom. The Floquet mode index $n,m$ in the Floquet Hamiltonian corresponds to the number of magnons arising from the time-periodic precession of the magnetization. Since the resonance frequency under FMR or AFMR is in the range of $1\,\mu {\rm eV}$ up to $10\,{\rm meV}$, the driving frequency $\Omega$ is much smaller than the band gap of TMDs (\textit{e.g.}, $1.7\,{\rm eV}$ for ${\rm MoS_2}$). The hybridization of the Floquet sidebands of the conduction band with those of the valence band is therefore strongly suppressed, one can hence safely neglect 
the effects of valence band under driving conditions and consider the driven dynamics of the conduction band electrons as described by the Floquet Hamiltonian Eq.~\eqref{widematrix}.
%
\begin{figure}
   \begin{center}
            \includegraphics[width=\columnwidth]{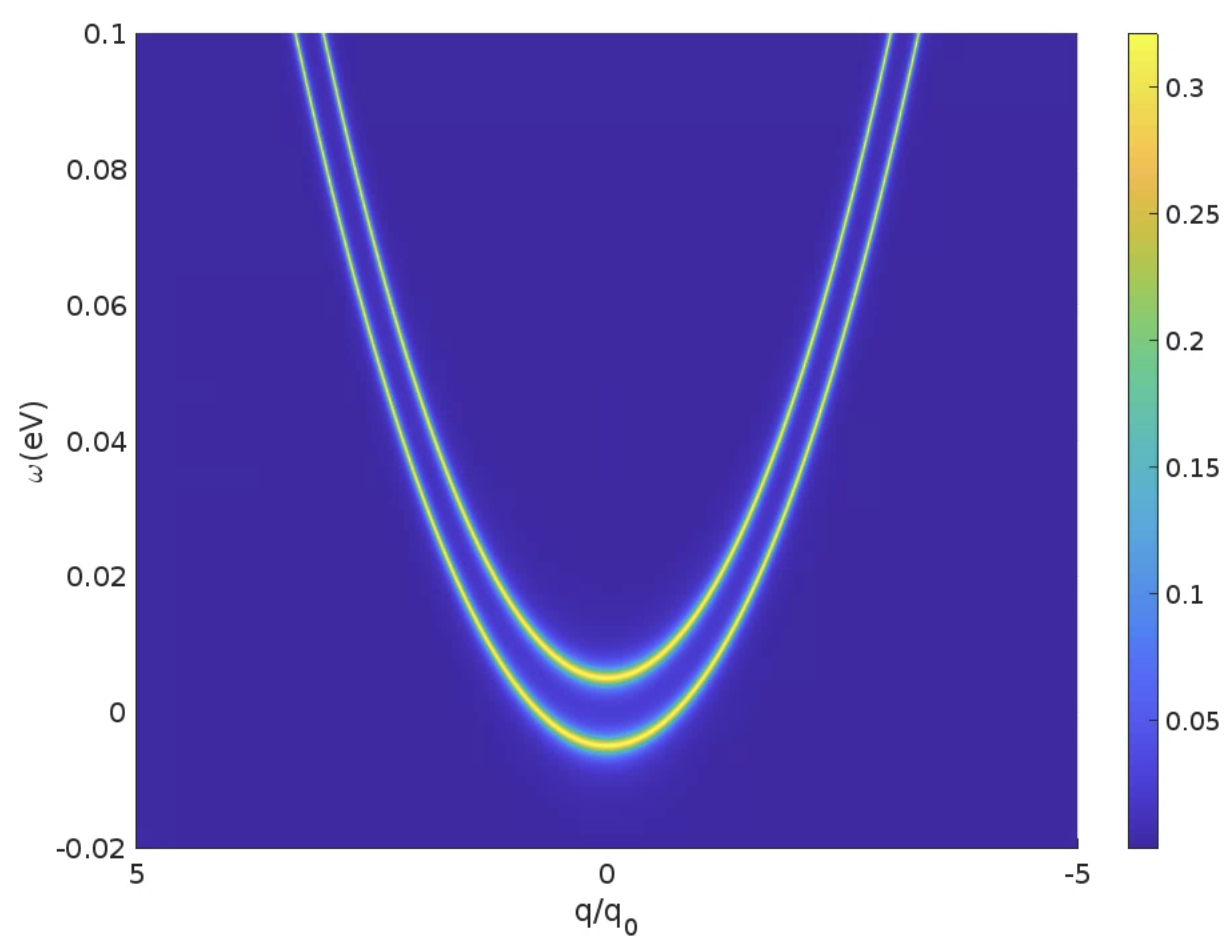}
                \end{center}
                \caption{Spectral function $\tilde{\mathcal{A}}_{\rm T,\tau\pm}(\omega)$  as a function of $q$ and extended zone frequency $\omega$ at a driving frequency $\hbar\Omega=10\,{\rm \mu eV}$, Fermi energy $\mu=10\,{\rm  meV}$, and exchange coupling strength $J_{\rm T}=5\,{\rm  meV}$.} \label{spectral}
\end{figure}

Observing that the above Floquet Hamiltonian Eq.~\eqref{widematrix} is a block diagonal matrix composing of $2\times2$ block matrices that mix two adjacent Floquet modes, we can write it in the following form of diagonal blocks~\cite{edge}:
\begin{eqnarray}
 &&H^{\rm T}_{F,\tau}=\bigoplus_{\nu\in\mathbb{Z}+\frac{1}{2}}H^{\rm T}_{\tau\nu},\\
 &&H^{\rm T}_{\tau\nu}=-\nu\Omega+\begin{bmatrix}
\varepsilon_{q\downarrow\tau}+\Omega/2&  J_{\rm T} \sin{\alpha}\\J_{\rm T} \sin{\alpha} &  \varepsilon_{q\uparrow\tau}-\Omega/2
 \end{bmatrix},\label{blockH}
\end{eqnarray}
and the corresponding retarded Green's function 
\begin{eqnarray} \label{GForig}
&&{\mathcal{G}}_{\rm T}^R=(\bar\omega+i\Gamma- H^{\rm T}_{F,\tau})^{-1}=\bigoplus_{\nu\in\mathbb{Z}+\frac{1}{2}} G_{{\rm T},\tau\nu}^R\\
&&=\bigoplus_{\nu\in\mathbb{Z}+\frac{1}{2}}(\bar\omega+i\Gamma- H^{\rm T}_{\tau\nu})^{-1}.\nonumber
 \end{eqnarray}
Here and henceforth, the Green's function's dependence on $\tau$ and $\nu$ are made explicit while its $q$-dependence is suppressed for compactness and clarity. Each block in Eq.~\ref{blockH} can be diagonalized by using the unitary transformation
\begin{eqnarray} \label{unit}
 U_{\rm T}=\begin{bmatrix}
\cos{\frac{\theta}{2}}&  \sin{\frac{\theta}{2}}\\-\sin{\frac{\theta}{2}}&  \cos{\frac{\theta}{2}}
 \end{bmatrix},
\end{eqnarray}
with $\tan{\theta}=2J_{\rm T} \sin{\alpha}/(\varepsilon_{q\uparrow\tau}-\varepsilon_{q\downarrow\tau}-\Omega)$. In the following, an overhead tilde is used to denote quantities in the diagonal basis. After transformation $\tilde H^{\rm T}_{\tau\nu}=U_{\rm T} H^{\rm T}_{\tau\nu} U_{\rm T}^\dagger$, we have the following $2\times2$ block of the top-layer Floquet Hamiltonian 
\begin{eqnarray} 
&&\tilde H^{\rm T}_{\tau\nu}=-\nu\Omega+\begin{bmatrix}
E^{\rm T}_{q,+}& 0\\0 &  E^{\rm T}_{q,-}
 \end{bmatrix}, \label{HTnu1} \\
&&E^{\rm T}_{q,\pm} =\frac{1}{2}(\varepsilon_{q\uparrow\tau}+\varepsilon_{q\downarrow\tau})\nonumber\\
&&\pm\frac{1}{2}\sqrt{(\varepsilon_{q\uparrow\tau}-\varepsilon_{q\downarrow\tau}-\Omega)^2+4J_{\rm T}^2\sin^2{\alpha}}, \label{HTnu2}
 \end{eqnarray}     
where $E^{\rm T}_{q,\pm}$ gives the band energies of the new quasiparticles obtained after diagonalization. 
To see what these quasiparticles actually are,   
let us denote the basis of the original spin-Floquet mode space for the Floquet Hamiltonian Eq.~\eqref{blockH} as $\cup^{\infty}_{n=-\infty}\{\phi_{\downarrow,n-1},\phi_{\uparrow,n}\}$. Then, the new basis after transformation is
\begin{eqnarray}
&&u_{\nu,+}=\cos{\frac{\theta}{2}}\phi_{\downarrow,n-1}+\sin{\frac{\theta}{2}}\phi_{\uparrow,n},\label{unup} \\
&&u_{\nu,-}=-\sin{\frac{\theta}{2}}\phi_{\downarrow,n-1}+\cos{\frac{\theta}{2}}\phi_{\uparrow,n}. \label{unum}
 \end{eqnarray}     
Thus 
the wave function of the new quasiparticles 
is a linear combination of the spin-up component of the $n^{\mathrm{th}}$ Floquet mode wave function with the spin-down component of the $(n-1)^{\mathrm{th}}$ Floquet mode wave function, and can aptly be called magnon-dressed electrons.  

%
%

Although the bottom layer is undriven, it is convenient to write its Hamiltonian in the Floquet representation as well, in order to treat it on an equal footing with the top layer. Its Floquet Hamiltonian takes the form of Eq.~\eqref{widematrix} with $J_{\rm T}=0$, which is diagonal and hence no unitary transformation is needed. The $2\times2$ blocks of the undriven bottom-layer Floquet Hamiltonian is
\begin{eqnarray}
&&\tilde H^{\rm B}_{\tau\nu}=-\nu\Omega+\begin{bmatrix}
E^{\rm B}_{q,+}& 0\\0 &  E^{\rm B}_{q,-}
 \end{bmatrix},\label{HBnu1}\\
&&E^{\rm B}_{q,-} =\varepsilon^0_{q\uparrow\tau}-\frac{1}{2}\Omega,~~~~~~E^{\rm B}_{q,+} =\varepsilon^0_{q\downarrow\tau}+\frac{1}{2}\Omega.\label{HBnu2}
 \end{eqnarray}     
%
Then, in the diagonal basis the retarded Floquet Green's function for the top and bottom layers are given by
\begin{eqnarray} \label{GFreduced}
&&\tilde{\mathcal{G}}_{\rm T/B}^R=(\bar\omega+i\Gamma-\tilde H^{\rm T/B}_\tau)^{-1}=\bigoplus_{\nu\in\mathbb{Z}+\frac{1}{2}}\tilde G_{{\rm T/B},\tau\nu}^R\\
&&=\bigoplus_{\nu\in\mathbb{Z}+\frac{1}{2}}(\bar\omega+i\Gamma-\tilde H^{\rm T/B}_{\tau\nu})^{-1},\nonumber
 \end{eqnarray}
with $\tilde H^{\rm T/B}_{\tau\nu}$ given by Eqs.~\eqref{HTnu1}-\eqref{HTnu2} and Eqs.~\eqref{HBnu1}-\eqref{HBnu2}, respectively. The  Green's function in the diagonal basis is related to the Green's function in the original basis by $\tilde G_{{\rm T},\tau\nu}^R=U_{\rm T}^\dagger G_{{\rm T},\tau\nu}^RU_{\rm T}$ for the top layer and $\tilde G_{{\rm B},\tau\nu}^R=G_{{\rm B},\tau\nu}^R$ for the bottom layer. The Floquet Green's functions in Eq.~\eqref{GFreduced} are defined in the reduced zone. In the rest of this paper,  
we will express all frequency-dependent quantities in terms of the physical, extended zone frequency $\omega=\bar\omega+\nu\Omega$. The extended-zone Green's functions in the diagonal basis are then given by \cite{Modi_1,Modi_2,asmar2022impurity} 
$\tilde G_{\rm T/B,\tau}^R(\omega)\equiv\tilde G_{{\rm T/B},\tau\ \nu=0}^R(\omega)$.  

As will be clear in the next section, it will be useful to define the diagonal matrix $\tilde{\mathcal{A}}_{\rm T,\tau}(\omega)= -(1/\pi)\textrm{Im} \tilde G^R_{\rm T,\tau}(\omega)$ with elements being the 
spectral functions for the new quasiparticles  labeled by $\beta\in\{+,-\}$,  
%
\begin{eqnarray}\label{spectralfcn}
\tilde{\mathcal{A}}_{\rm T,\tau\beta}(\omega)&=& -\frac{1}{\pi}\textrm{Im} \tilde G^R_{\rm T,\tau\beta}(\omega) \nonumber \\
&=& \frac{\Gamma}{\pi}\tilde G^R_{\rm T,\tau\beta}(\omega)\tilde G^A_{\rm T,\tau\beta}(\omega), 
\end{eqnarray}
where $\tilde G^R_{{\rm T},\tau\beta}\equiv [\tilde G^R_{{\rm T},\tau}(\omega)]_{\beta\beta}$ is the diagonal element of the Green's function $\tilde G^R_{{\rm T},\tau}(\omega)$. Fig.~\ref{spectral}  illustrates the spectral functions for the top layer in the extended zone. Definitions of $\tilde{\mathcal{A}}_{\rm B,\tau}$ for the bottom layer can be made similarly as the above.

\section{Tunneling Spin and Charge Currents}\label{currentsec}

Having laid out the Floquet Green's functions for both layers, in this section we formulate the nonequilibrium tunneling transport problem and derive the expressions of the tunneling spin and charge currents in terms of the Floquet Green's functions.

The current flowing between the two layers can be calculated by considering the change of the total number of electron charges and spins of a single layer \cite{Scalapino}. The total charge operator and spin operator in the bottom layer are given by $\hat Q=\sum_{qs\tau}\hat{c}^{\dagger}_{qs\tau}\hat{c}_{qs\tau}$ and $\hat{S_z}=\sum_{qss'\tau}\hat{c}^{\dagger}_{qs\tau}[\sigma_z]_{ss'}\hat{c}_{qs'\tau}$, where $[\sigma_z]_{ss'}$ is the z-component of the Pauli matrices acting on the spin indices. We introduce the spin current with the same units as charge current defined by $I_{S_z}(t)=-(2e/\hbar) \langle d\hat{S}_z/dt\rangle$. Using the Heisenberg equation of motion to calculate the change of the total spin operator and the charge operator in the bottom layer, the corresponding spin current in terms of the tunneling Hamiltonian is
\begin{eqnarray}
I_{S_z}(t)=-e\frac{i}{\hbar}\langle[H_{\rm I},\hat{S}_z]\rangle,
\end{eqnarray}
due to the fact that the commutators with other terms   $[H_{\rm T},\hat{S}_z]$ and also $[H_{\rm B},\hat{S}_z]$ are zero. Similarly, because the commutators  $[H_{\rm T},\hat Q]$ and $[H_{\rm B},\hat Q]$ are zero, the charge current is 
\begin{eqnarray}
I_{\rm C}(t)=-e\frac{i}{\hbar}\langle[H_{\rm I},\hat Q]\rangle. 
\end{eqnarray}

Therefore, by evaluating the commutators, the spin current can be expressed in terms of the Green's function as 
%
\begin{eqnarray} \label{ISz1}
I_{S_z}(t)=-\frac{e}{\hbar}{\rm Tr}\sum_{\tau\tau'}\sum_{q}\left\{V^*G^<_{\tau',\tau}(t,t)\sigma_z\right.\nonumber\\
\left.-VG^<_{\tau,\tau'}(t,t)\sigma_z\right\},
\end{eqnarray}
where the trace is taken over the spin degree of freedom and $G^<_{\tau,\tau'}(t,t')$ is the $2\times2$ interacting lesser Green's function coupling the two layers with its components defined by $[G^<_{\tau,\tau'}(t,t')]_{ss'} = i\langle\hat{c}^{\dagger}_{qs'\tau'}(t')\hat{d}_{qs\tau}(t)\rangle$.
This interacting Green's function can be written in terms of the non-interacting Green's functions of the individual layers following Ref.~\cite{Jahuo}. 
Taking the time derivative of the interacting contour-ordered Green's function as defined above allows us to obtain its equation of motion, which can be inverted to yield the interacting Green's function in terms of the noninteracting single-layer Green's functions. Eq.~\eqref{ISz1} can then be written as
\begin{eqnarray}
&&I_{S_z}(t)=\frac{2e}{\hbar}{\rm Re}\bigg[{\rm Tr}\sum_{\tau\tau'}\sum_{q}VG^<_{\tau,\tau'}(t,t)\sigma_z\bigg]\\
&&=\frac{2e}{\hbar}{\rm Re}\bigg\{{\rm Tr}\sum_{\tau\tau'q}\int dt_1\big[G^R_{{\rm T},\tau}(t,t_1)G^<_{{\rm B},\tau'}(t_1,t)\sigma_z\nonumber\\
&&+G^<_{{\rm T},\tau}(t,t_1)G^A_{{\rm B},\tau'}(t_1,t)\sigma_z\big]|V|^2\bigg\}.\nonumber
\end{eqnarray}
Then, using the fact that the bottom-layer Green's functions are diagonal in the spin space, we do time averaging and go into the Floquet representation. By summing over all the matrices of different $\nu$, transitioning into extended zone frequency and making the change of variable $\omega=\bar\omega+\nu\Omega$ we can combine them into an integral of extended zone frequency $\omega\in (-\infty,\infty) $. The time-averaged tunneling spin current can then be written as 
\begin{eqnarray} \label{SzEq1}
&&I_{S_z}=\frac{e|V|^2}{2\pi\hbar} {\rm Tr}\sum_{q \tau\tau'}\int _{-\infty}^{\infty}d\omega \{\big[G^R_{{\rm T},\tau}(\omega)-G^A_{{\rm T},\tau}(\omega)\big]\nonumber\\
&&G^<_{{\rm B},\tau'}(\omega)\sigma_z-G^<_{{\rm T},\tau}(\omega)\big[G^R_{{\rm B},\tau'}(\omega)-G^A_{{\rm B},\tau'}(\omega)\big]\sigma_z\},\nonumber\\
\end{eqnarray}
where the lesser Green's function $\tilde G^<(\omega)$ is given by Eq.~\eqref{lesserG}.  

Following a similar line, the expression for the time-averaged tunneling charge current can be obtained as
\begin{eqnarray} \label{CEq1}
&&I_{\rm C}=\frac{e|V|^2}{2\pi\hbar} {\rm Tr}\sum_{q \tau\tau'}\int _{-\infty}^{\infty}d\omega \{\big[G^R_{{\rm T},\tau}(\omega)-G^A_{{\rm T},\tau}(\omega)\big]\nonumber\\
&&G^<_{{\rm B},\tau'}(\omega)-G^<_{{\rm T},\tau}(\omega)\big[G^R_{{\rm B},\tau'}(\omega)-G^A_{{\rm B},\tau'}(\omega)\big]\}.\nonumber\\
\end{eqnarray}

The above results simplify when the Green's functions are expressed in the diagonal basis Eqs.~\eqref{unup}-\eqref{unum} using the unitary transformation Eq.~\eqref{unit}.
Then, for both layers, using the definitions of $\tilde{\mathcal{A}}_{\tau}$ in Eqs.~\eqref{spectralfcn} 
and the expression of the lesser Green's function ${G}_{\tau}^<$ 
in Eq.~\eqref{lesserG}, Eq.~\eqref{SzEq1} can be written as  
\begin{eqnarray}\label{SzEq2}
&&I_{S_z}=\frac{2\Gamma e}{\hbar}|V|^2 \sum_{q\tau\tau'}\int_{-\infty}^{\infty}d\omega \\
&& \textrm{Tr}\{U_{\rm T}^\dagger\tilde{\mathcal A}_{\rm T,\tau}(\omega)U_{\rm T}\tilde G^R_{\rm B,\tau'}(\omega)F_{\rm B,\tau'}(\omega)\tilde G^A_{\rm B,\tau'}(\omega)\sigma_z\nonumber\\
&&- \tilde{\mathcal A}_{\rm B,\tau'}(\omega)U_{\rm T}^\dagger\tilde G^R_{\rm T,\tau}(\omega)U_{\rm T}F_{\rm T,\tau}(\omega)U_{\rm T}^\dagger\tilde G^A_{\rm T,\tau}(\omega)U_{\rm T}\sigma_z\}\nonumber,
\end{eqnarray}
where 
\begin{eqnarray}
&&F_{\rm T,B}(\omega)=\begin{bmatrix}
f_{\rm T,B}(\omega-\frac{1}{2}\Omega)& 0\\0 &  f_{\rm T,B}(\omega+\frac{1}{2}\Omega)
 \end{bmatrix},
 \end{eqnarray}     
and $f_{\rm T,B}(\omega)=1/[e^{(\omega-\mu_{\rm T,B})/k_BT}+1]$ is the Fermi distribution of the top (T) and bottom (B) layers having chemical potentials $\mu_{\rm T}$ and $\mu_{\rm B}$, respectively. Simplifying Eq.~\eqref{SzEq1}, we obtain our final expression of the time-averaged spin current as 

\begin{eqnarray}\label{Is}
&&I_{S_z}=-\frac{e\pi}{\hbar}|V|^2 \sum_{q\tau\tau'}\sum_{\gamma\beta}\int_{-\infty}^{\infty} d\omega\,\gamma\bigg\{
(1+\gamma\beta\cos{\theta})\nonumber\\
&&\times\tilde{\mathcal A}_{\rm T,\tau\beta}(\omega)\tilde{\mathcal A}_{\rm B,\tau'\gamma}(\omega)\left[f_{\rm T}(\omega+\gamma\frac{\Omega}{2})-f_{\rm B}(\omega+\gamma\frac{\Omega}{2})\right]\nonumber\\
&&-\frac{\pi}{\Gamma}J_T^2 \sin^2{\alpha}\ \tilde{\mathcal A}_{\rm T,\tau+}(\omega)\tilde{\mathcal A}_{\rm T,\tau-}(\omega)\tilde{\mathcal A}_{\rm B, \tau'\gamma}(\omega)\nonumber\\
&&\times\left[f_{\rm T}(\omega+\gamma\frac{\Omega}{2})-f_{\rm T}(\omega-\gamma\frac{\Omega}{2})\right]\bigg\},
\end{eqnarray}

where $\gamma,\beta \in \{+,-\}$.

Similarly, the time-averaged charge current follows from Eq.~\eqref{CEq1}   as 

\begin{eqnarray}\label{Ic}
&&I_{\rm C}=-\frac{e\pi}{\hbar}|V|^2 \sum_{q\tau\tau'}\sum_{\gamma\beta}\int_{-\infty}^{\infty} d\omega \bigg\{(1+\gamma\beta\cos{\theta})\nonumber\\
&&\times\tilde{\mathcal A}_{\rm T,\tau\beta}(\omega)\tilde{\mathcal A}_{\rm B,\tau'\gamma}(\omega)\left[f_{\rm T}(\omega+\gamma\frac{\Omega}{2})-f_{\rm B}(\omega+\gamma\frac{\Omega}{2})\right]\nonumber\\
&&-\frac{\pi}{\Gamma}J_T^2 \sin^2{\alpha}\ \tilde{\mathcal A}_{\rm T,\tau+}(\omega)\tilde{\mathcal A}_{\rm T,\tau-}(\omega)\tilde{\mathcal A}_{\rm B, \tau'\gamma}(\omega)\nonumber\\
&&\times\left[f_{\rm T}(\omega+\gamma\frac{\Omega}{2})-f_{\rm T}(\omega-\gamma\frac{\Omega}{2})\right]\bigg\}.
\end{eqnarray}

Eqs.~\eqref{Is}-\eqref{Ic} are the main results in this section. Each of them contains two contributions, the first one depending on the form factor $(1+\gamma\beta\cos{\theta})$ and the second one on $\sin^2{\theta}$. In the absence of spin pumping where the magnetization is stationary, $\alpha = \theta = 0$, the second contribution in each of Eqs.~\eqref{Is}-\eqref{Ic} vanishes. In the spin current Eq.~\eqref{Is}, the remaining term given by the first contribution is non-zero only when $\gamma = \beta$, and terms cancel pairwise under the remaining sum over $\gamma,\tau,\tau'$ to give the physically expected result of a zero spin current in the absence of a precessing magnetization, regardless of whether  $\mu_{\rm T} =  \mu_{\rm B}$ or not. For the charge current, the first contribution is reminiscent of the conventional formula of the tunneling charge current arising from unbalanced chemical potentials. This contribution is present whenever $\mu_{\rm T} \ne \mu_{\rm B}$, either when the system is driven or undriven. 

On the contrary, when the two layers have the same chemical potentials with $\mu_{\rm T} = \mu_{\rm B}$, the first contribution in each of Eqs.~\eqref{Is}-\eqref{Ic} drops out. Therefore, we can see that the second contribution is not driven by an unbalanced chemical potential but is purely driven by the precessing magnetization. In the next section, we shall focus our attention on this contribution, which is unique to our magnetization-driven case. 

Before moving on to the next section, we remark on the presence of a vertical charge current given by the second contribution in Eq.~\eqref{Ic} even when $\mu_{\rm T} = \mu_{\rm B}$. To appreciate why that is the case, 
we briefly digress for a moment to consider the slightly more general scenario when the two layers, with  chemical potentials $\mu_{\rm T}$ and $\mu_{\rm B}$, are both coupled to a precessing magnetization driven at the same frequency. 
We find that the corresponding time-averaged charge current is given by the expression  

\begin{eqnarray}\label{current4}
&&I_{\rm C}=-\frac{e\pi}{2\hbar}|V|^2 \sum_{q\tau\tau'}\sum_{\gamma\beta S}\int_{-\infty}^{\infty} d\omega \bigg\{\tilde{\mathcal A}_{\rm T,\tau\beta}(\omega)\tilde{\mathcal A}_{\rm B,\tau'\gamma}(\omega)\nonumber\\
&&\times\bigg[[1+\gamma\beta\cos{(\theta_{\rm T}-\theta_{\rm B})}][f_{\rm T}(\omega+S\frac{\Omega}{2})(1+\beta S\cos{\theta_{\rm T}})\nonumber\\
&&-f_{\rm B}(\omega+S\frac{\Omega}{2})(1+\gamma S\cos{\theta_{\rm B}})]+\gamma S\sin{(\theta_{\rm T}-\theta_{\rm B})}\nonumber\\
&&\times[\sin{\theta_{\rm B}}f_{\rm B}(\omega+S\frac{\Omega}{2})+\sin{\theta_{\rm T}}f_{\rm T}(\omega+S\frac{\Omega}{2})]\bigg]\nonumber\\
&&-2\frac{\pi}{\Gamma}\sin^2{\alpha}\sin{(\theta_{\rm T}-\theta_{\rm B})}\gamma S\bigg[\tilde{\mathcal A}_{\rm B,\tau'+}(\omega)\tilde{\mathcal A}_{\rm B,\tau'-}(\omega)\tilde{\mathcal A}_{\rm T,\tau\gamma}(\omega) \nonumber\\
&&\times\frac{J_B^2}{\sin{\theta_{\rm B}}}f_{\rm B}(\omega+S\frac{\Omega}{2})+\tilde{\mathcal A}_{\rm T,\tau+}(\omega)\tilde{\mathcal A}_{\rm T,\tau-}(\omega)\tilde{\mathcal A}_{\rm B,\tau'\gamma}(\omega)\frac{J_T^2}{\sin{\theta_{\rm T}}}\nonumber\\
&&\times f_{\rm T}(\omega+S\frac{\Omega}{2})\bigg]\bigg\}. 
\end{eqnarray}

%
One can easily verify that this equation recovers  Eq.~\eqref{Ic} if we set  $\theta_{\rm B}=0$, corresponding to an undriven bottom layer (See the Appendix). Now, when both layers have the same chemical potentials $\mu_{\rm T} = \mu_{\rm B}$ and are driven by the same precessing magnetization with $J_{\rm T}=J_{\rm B}$, then $\theta_{\rm T}=\theta_{\rm B}$ from the definition of $\theta_{\rm T/B}$ below Eq.~\ref{unit}. In such case the heterostructure is completely top-down symmetric, and it can be seen that the right-hand side of Eq.~\eqref{current4} gives zero. Therefore, we can see that a nonzero charge current as obtained in Eq.~\eqref{Ic} only arises when one of the layers is driven, which results in a breaking of the top-down symmetry of the heterostructure.

\section{Results and Discussions} \label{numerics}

In this section we present our numerical and analytic results obtained from the main equations Eqs.~\eqref{Is}-\eqref{Ic}. In order to focus our attention on the unique contribution purely due to spin pumping, in what follows we take both layers to have equal chemical potentials so that the conventional contribution  [$\sim (1+\gamma \beta \cos\theta)$ in Eqs.~\eqref{Is}-\eqref{Ic}] due to interlayer bias drops out. 
In our numerical calculations below, we take the TMD parameters of ${\rm MoS_2}$ with $\Delta=1.7\, {\rm eV}$, $v=5\times10^5 \,{\rm m/s}$, and $\lambda=75\,{\rm meV}$~\cite{TMD,TMD2}. 

\subsection{Adiabatic Regime} \label{adia}

\begin{figure}
   \begin{center}
            \includegraphics[width=\columnwidth]{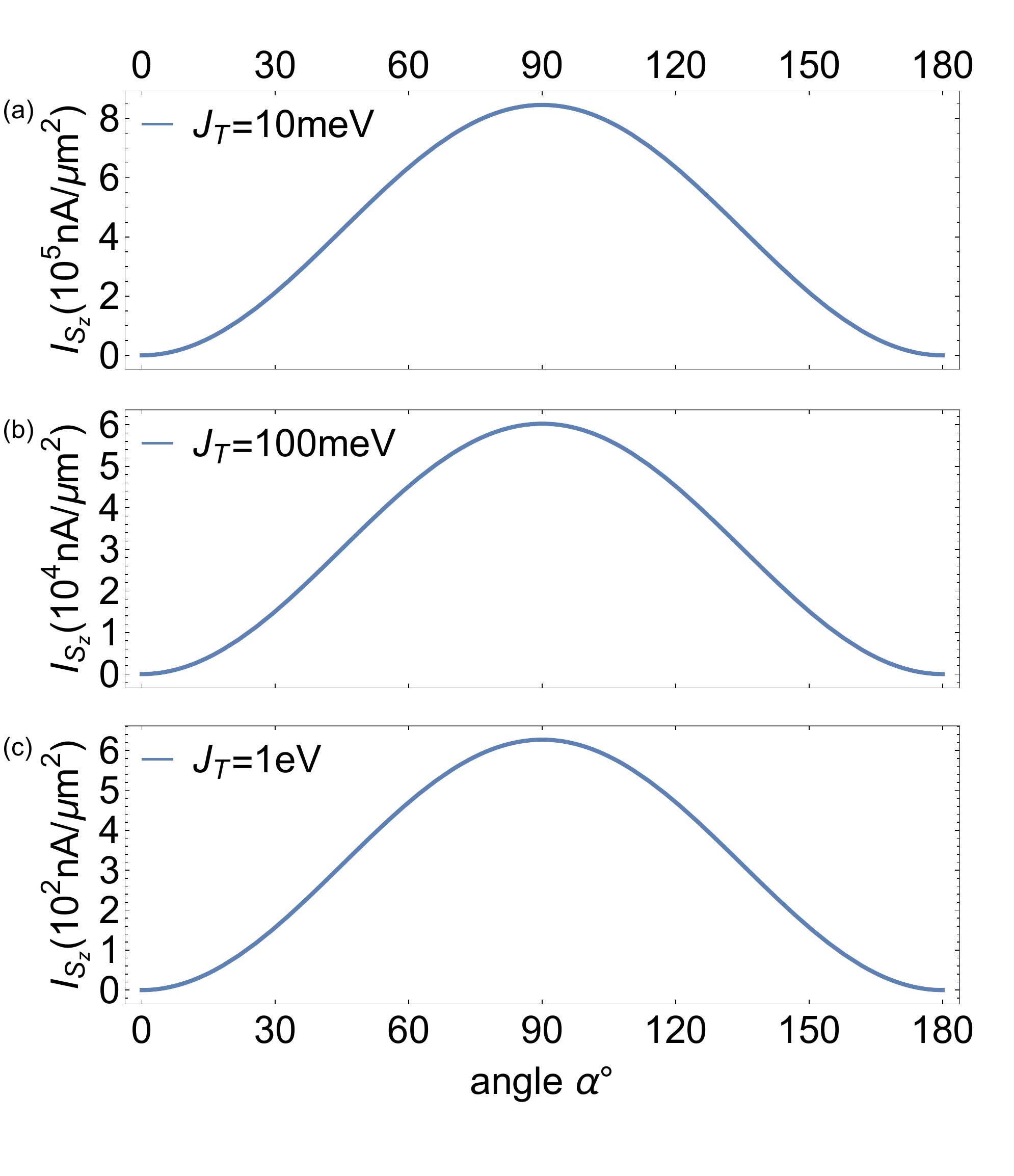}
                \end{center}
                \caption{Time-averaged tunneling spin current $I_{S_z}$ as a function of precession angle $\alpha$ with the top layer driven at frequency $\hbar\Omega=1\,{\rm \mu eV}$, where the top and bottom layers have the same Fermi energy $\mu=10\,{\rm  meV}$. The top-layer coupling strength is $J_{\rm T}=10, 100\, {\rm  meV}$ and $1 {\rm  eV}$ , $\Gamma = 10 {\rm meV}$ and tunneling amplitude $V= 36\, {\rm meV}$.} \label{3}
\end{figure}

\begin{figure}
   \begin{center}
            \includegraphics[width=\columnwidth]{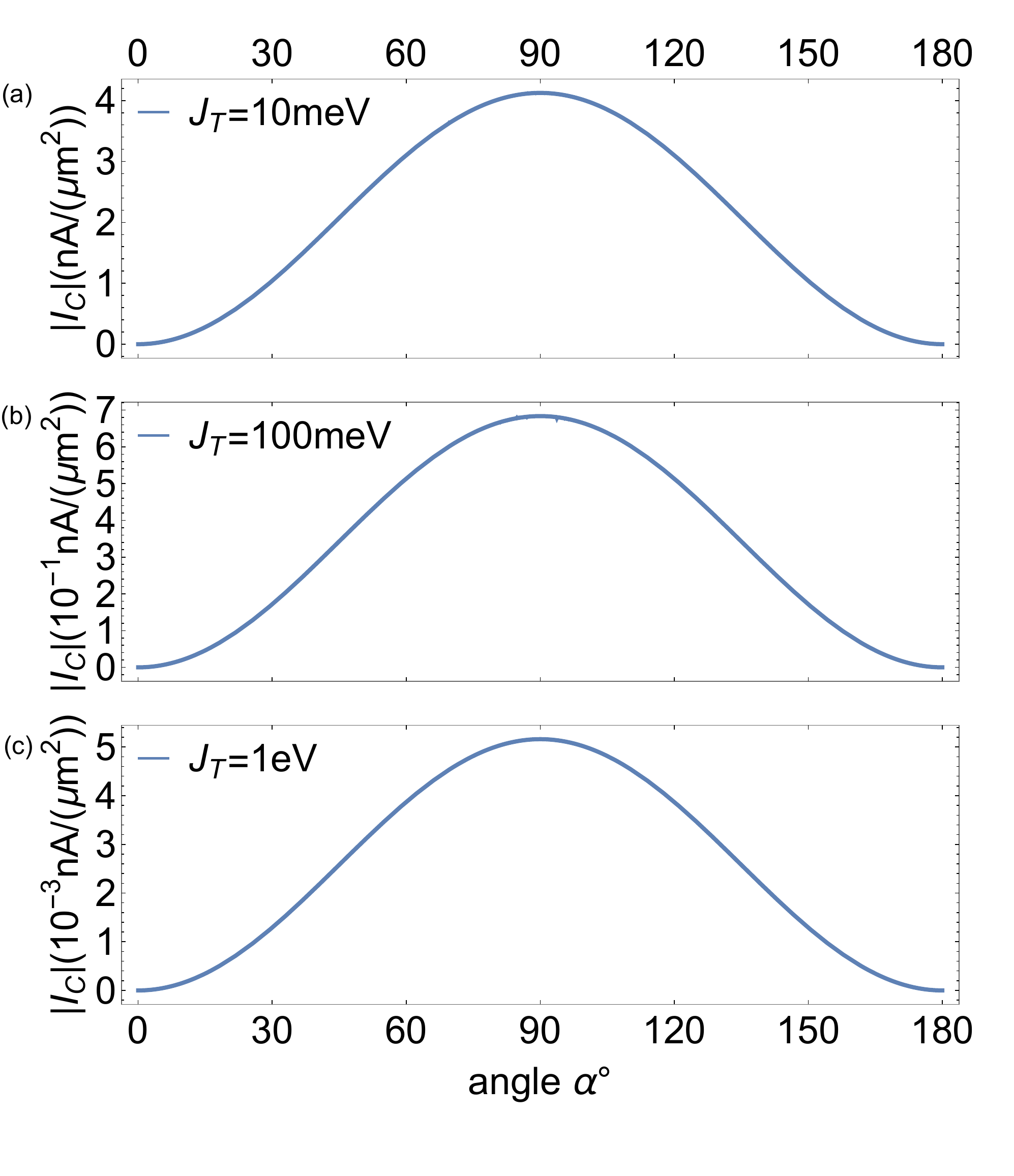}
                \end{center}
                \caption{Time-averaged tunneling charge current $|I_C|$ as a function of precession angle $\alpha$. The parameters $J_{\rm T},\Gamma, V,\mu$ and $\hbar\Omega$ are the same as in Fig.~\ref{3}.}\label{9}

\end{figure}

We first focus on the adiabatic regime where $\hbar \Omega/J_{\rm T} \ll 1$. To see how the currents vary with the precession angle $\alpha$ of the magnetization vector, Fig.~\ref{3} shows the time-averaged tunneling spin current $I_{S_z}$ as a function of $\alpha$ at a  fixed frequency $\hbar \Omega = 1\,\mu\mathrm{eV}$ for different values of exchange coupling $J_{\rm T} = 10\,\mathrm{meV}, 100\,\mathrm{meV}, 1\,\mathrm{eV}$. In all these cases, we find that the spin current can numerically be well fitted by a  $\sin^2{\alpha}$ dependence. We note that the same dependence is also reported in the spin current pumped by a single precessing spin through a one-dimensional tight-binding toy model~\cite{Nick1}. We also find the same precession angle dependence $\sin^2{\alpha}$ for the charge current $I_C$ as shown in Fig.~\ref{9}. For both spin and charge currents, their magnitudes are found to decrease with increasing exchange coupling $J_{\rm T}$ at the considered driving frequency. 

\begin{figure}
   \begin{center}
            \includegraphics[width=\columnwidth]{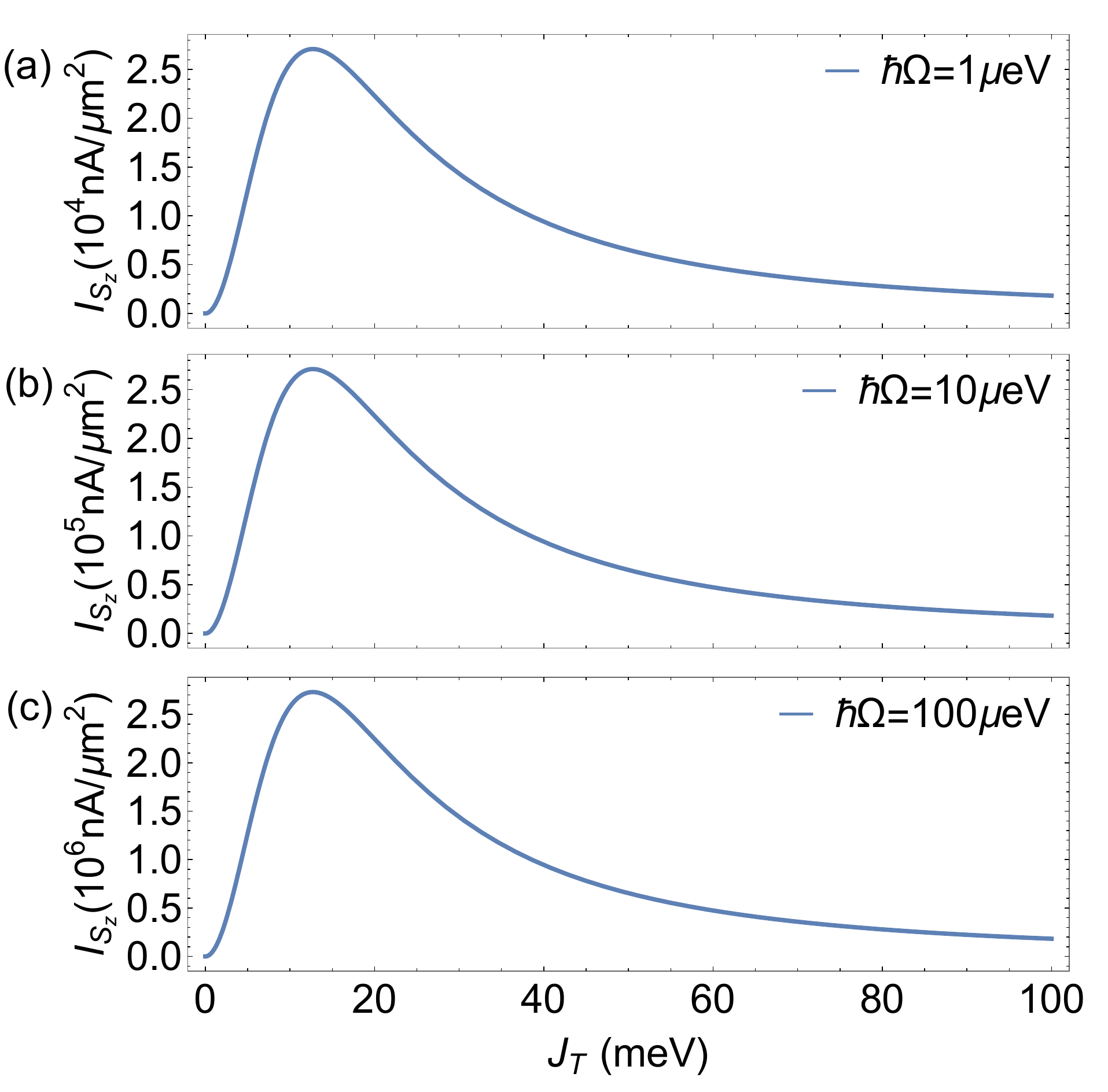}
                \end{center}
                \caption{Time-averaged tunneling spin current $I_{S_z}$ as a function of exchange coupling $J_{\rm T}$   at precession angle $\alpha=10\degree$. The driving frequency is $\hbar\Omega=1,10, 100\,{\rm \mu eV}$. The other parameters $\Gamma, V,\mu$  are the same as in Fig.~\ref{3}.} \label{IS_J}
\end{figure}

\begin{figure}
   \begin{center}
            \includegraphics[width=\columnwidth]{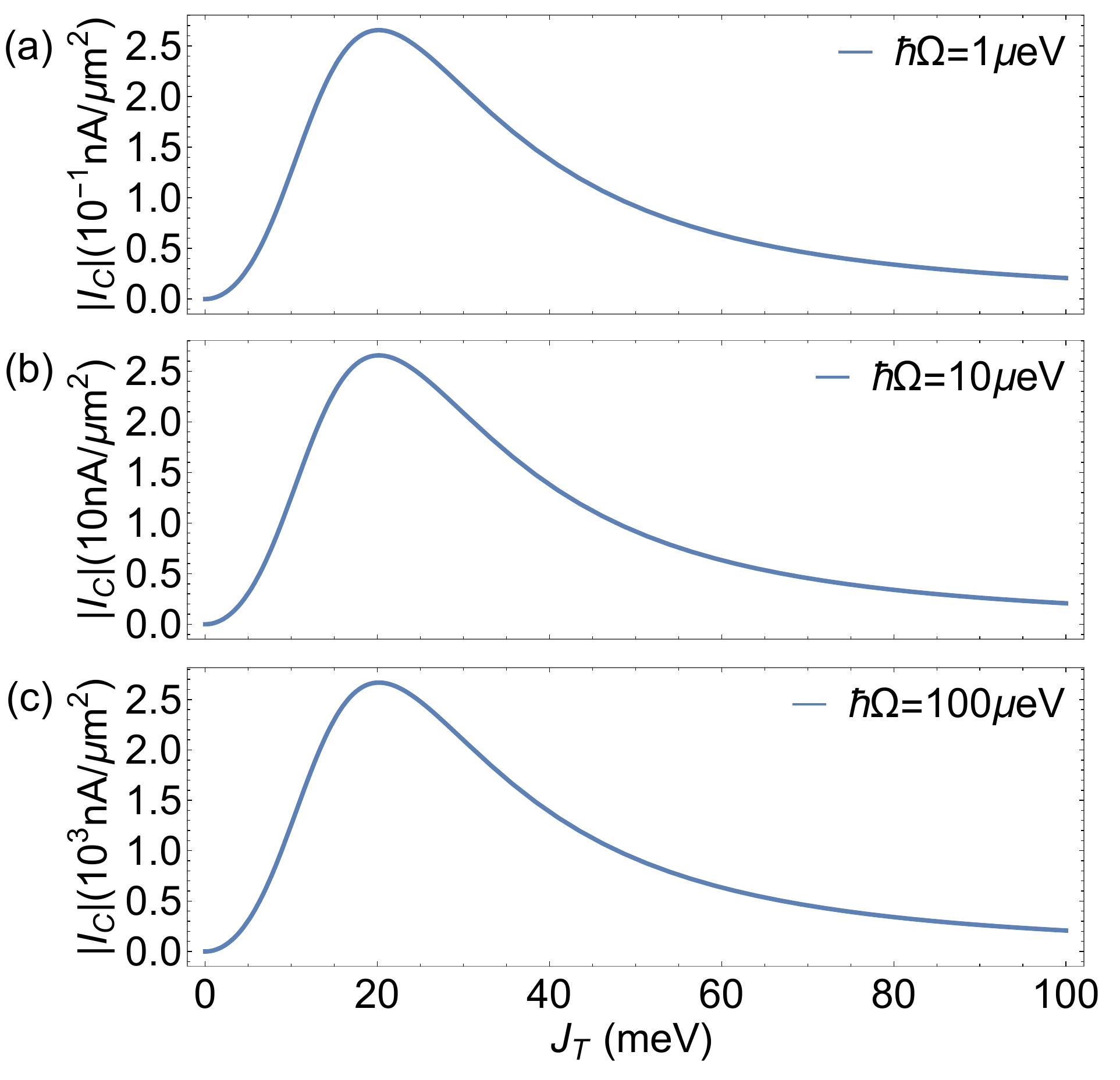}
                \end{center}
                \caption{Time-averaged tunneling charge current $|I_C|$ as a function of exchange coupling $J_{\rm T}$ 
                at precession angle $\alpha=10\degree$.  The driving frequency is $\hbar\Omega=1,10, 100\,{\rm \mu eV}$. The other parameters $\Gamma, V,\mu$ are the same as in Fig.~\ref{3}.}\label{IC_J}
\end{figure}

To gain a more complete picture on how both currents vary with $J_{\rm T}$, Figs.~\ref{IS_J}-~\ref{IC_J} show $I_{S_z}$ and $I_C$ as a function of $J_{\rm T}$ for different values of the driving frequency $\hbar \Omega = 1\,\mu\mathrm{eV}, 10\,\mu\mathrm{eV}, 100\,\mu\mathrm{eV}$. One can now observe from Fig.~\ref{IS_J} that $I_{S_z}$ first increases with $J_{\rm T}$ towards a maximum,  before decreasing gradually. The same behavior is displayed by $I_C$ in Fig.~\ref{IC_J}, albeit reaching the maximum at a different $J_{\rm T}$. 
Although $I_C$ behaves similarly as $I_{S_z}$ as a function of $\alpha$ and $J_{\rm T}$, $I_C$ is about $3\,-\,5$ orders of magnitude smaller than $I_{S_z}$. As discussed in Sec.~\ref{currentsec}, there is a finite tunneling charge current only  when the top-bottom symmetry of the heterostructure is broken by the precessing magnetization coupled to the top layer, an effect also reported in Ref.~\cite{Nick1}.  
%

\begin{figure}
   \begin{center}
            \includegraphics[width=\columnwidth]{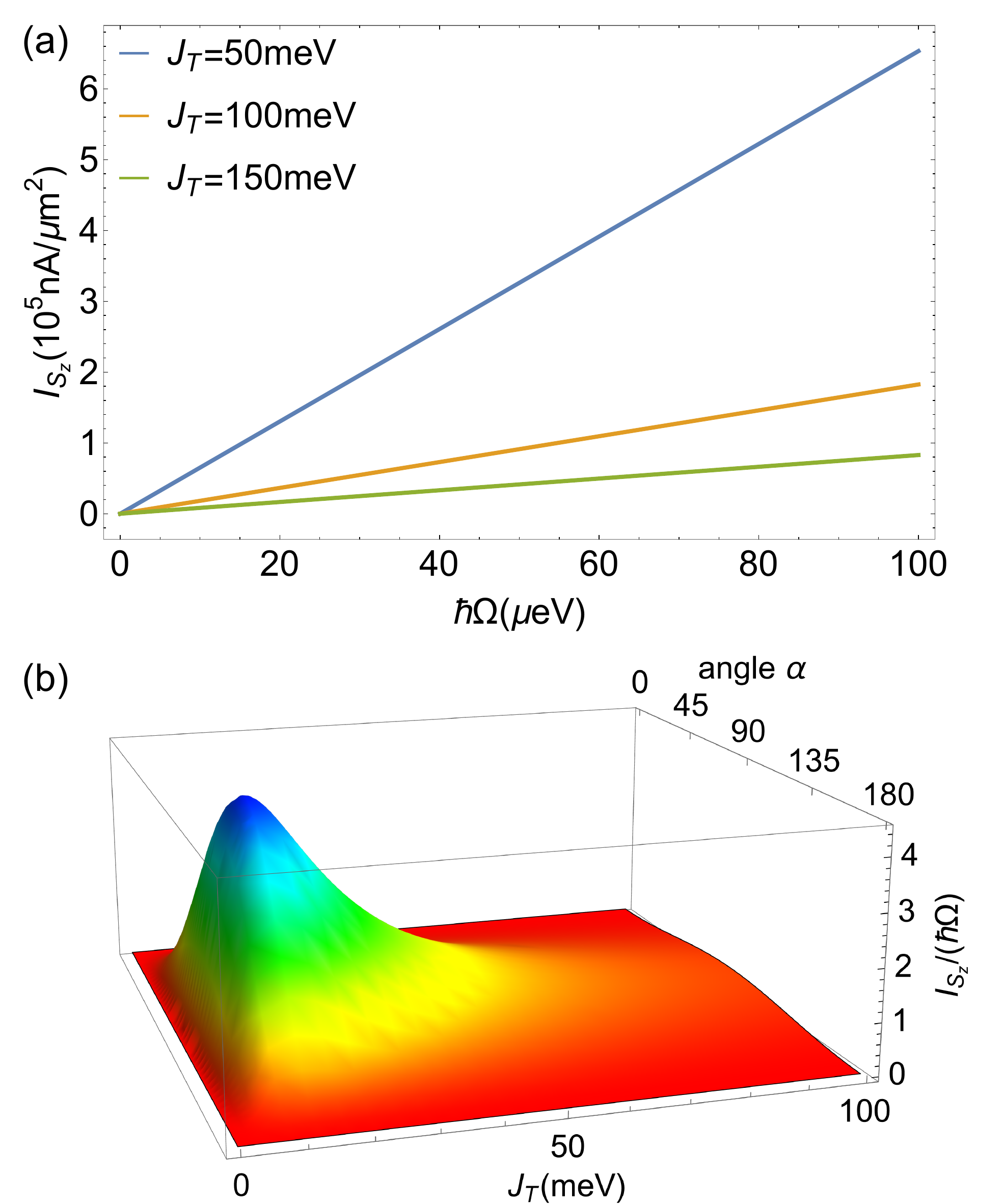}
                \end{center}
                \caption{Time-averaged tunneling spin current (a) $I_{S_z}$  as a function of driving frequency within the microwave frequency range at   precession angle $\alpha=10\degree$ and (b) $I_{S_z}/\hbar \Omega$ as a function of precession angle $\alpha$ and exchange coupling  $J_{\rm T}$. The other parameters $\Gamma, V,\mu$ and $\hbar\Omega$ 
 are the same as in Fig.~\ref{3}.} \label{5}
\end{figure}

We then examine the relation of the spin current with the driving frequency. Fig.~\ref{5}(a) shows the $I_{S_z}$ as a function of $\hbar \Omega$ up to $100\,\mu\mathrm{eV}$ for different values of $J_{\rm T}$. Note that these frequency values are still within the microwave frequency range. The plot displays a linearly increasing behavior with $\hbar \Omega$, consistent with the well-known linear relationship between the pumped spin current and the driving frequency within the adiabatic pumping regime~\cite{Nick1,pump2}. This is expected since the adiabatic condition $\hbar \Omega \ll J_{\rm T}$ is still satisfied by the parameters in Fig.~\ref{5}(a). Similar to the rotating frame picture \cite{revpump,Nick1}, $\hbar \Omega$ in the Floquet picture plays the role of an  ``interlayer bias voltage'', and thus the quantity $I_{S_z}/(\hbar \Omega)$ can be regarded as an effective  ``tunneling spin conductance'' appropriate for the case of spin pumping. It is a constant independent of frequency within the adiabatic regime. Fig.~\ref{5}(b) shows how this effective spin conductance varies with exchange coupling $J_{\rm T}$ and precession angle $\alpha$.  


%

To complement our numerical results, here we also provide approximate analytical results valid in the adiabatic regime. Again, we take the top and bottom layers to have the same chemical potentials and focus only on the second, solely precession-driven contribution in Eq.~\eqref{Is}. Expanding Eq.~\eqref{Is} up to leading order in frequency $\Omega$ and to linear order in the spin-orbit coupling strength $\lambda/\Delta$, we find that the leading-order result for the spin current is indeed linear in $\Omega$ as given by 
\begin{eqnarray}\label{IsEX}
&&I_{S_z}=\frac{8\Gamma e|V|^2m\Omega \sin^2{\alpha}}{\hbar^3(J_{\rm T}^2+\Gamma^2)(J_{\rm T}^2+4\Gamma^2)}\bigg[3\pi J_{\rm T}^2\nonumber\\
&&-(J_{\rm T}^2-2\Gamma^2)\cot^{-1}{\frac{\Gamma}{J_{\rm T}-\mu}}+4(J_{\rm T}^2+\Gamma^2)\cot^{-1}{\frac{\Gamma}{\mu}}\\
&&+(J_{\rm T}^2-2\Gamma^2)\cot^{-1}{\frac{\Gamma}{J_{\rm T}+\mu}}+3J_{\rm T}\Gamma\tanh^{-1}{\frac{2J_{\rm T}\mu}{J_{\rm T}^2+\Gamma^2+\mu^2}}\bigg]. \nonumber
\end{eqnarray}
In deriving the above, we find that the first order term in $\lambda/\Delta$ drops out, and therefore the spin current does not depend on the spin-orbit coupling strength up to linear order. 

\begin{figure}
   \begin{center}
            \includegraphics[width=\columnwidth]{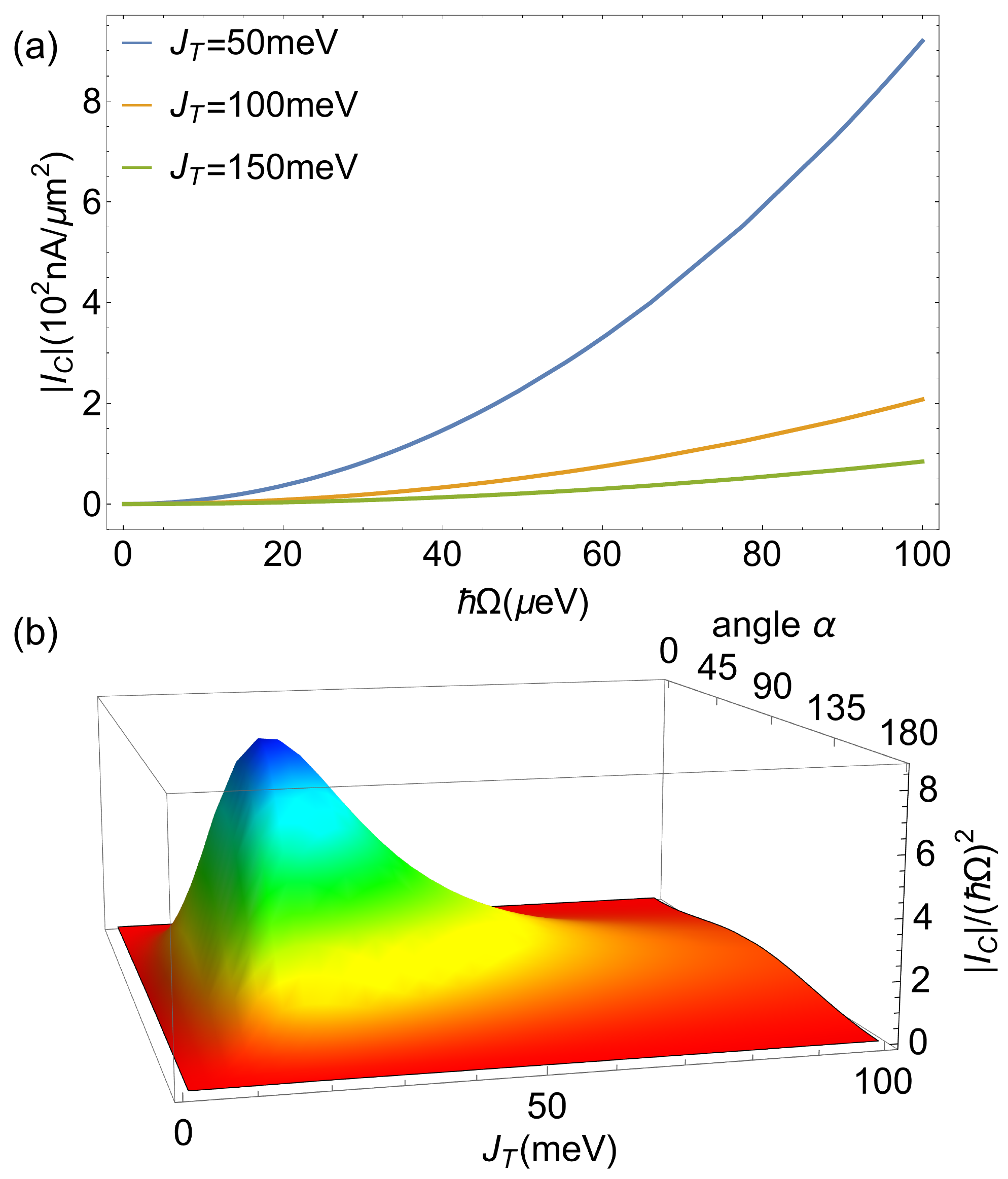}
                \end{center}
                \caption{Time-averaged tunneling charge current (a) $|I_C|$  as a function of driving frequency at  precession angle $\alpha=10\degree$ and (b) $|I_C|/(\hbar \Omega)^2$ as a function of precession angle $\alpha$ and exchange coupling $J_{\rm T}$. The other parameters $\Gamma, V,\mu$  are the same as in Fig.~\ref{3}.} \label{11}
\end{figure}

\begin{figure}
   \begin{center}
            \includegraphics[width=\columnwidth]{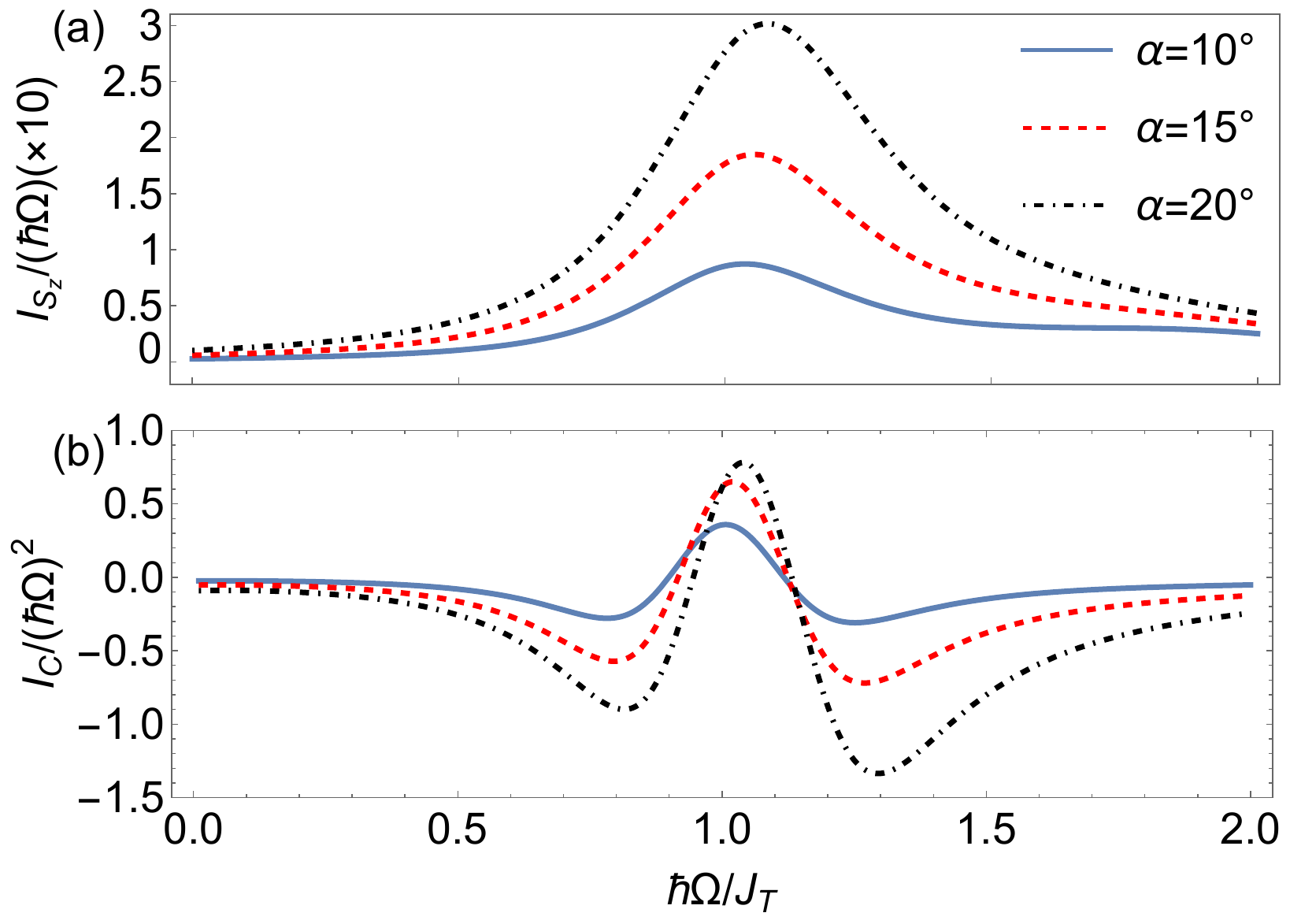}
                \end{center}
                \caption{Linear-log plot of time-averaged tunneling current  $I_{S_z}/\hbar \Omega$  and $|I_C|/(\hbar \Omega)^2$ as a function of  $\hbar \Omega/J_{\rm T}$ at precession angles $\alpha=10\degree, 15\degree, 20\degree$. The exchange coupling  is $J_{\rm T}=10\,{\rm  meV}$, $\Gamma=1{\rm mev}$ and the other parameters $ V,\mu$ are the same as in Fig.~\ref{3}.}\label{ILinLin}
\end{figure}

As a function of the same range of frequencies, Fig.~\ref{11}(a) shows the tunneling charge current $I_C$ for different values of $J_{\rm T}$. A plot of 
$I_C/(\hbar \Omega)^2$ versus $\hbar \Omega$ shows that the plot is also flat (see Fig.~\ref{ILinLin} in Sec.~\ref{nonadia}), indicating that 
$I_C$ displays a quadratic frequency dependence in the adiabatic pumping regime, consistent with Refs.~\cite{Nick3,Foa,Vavilov}. Using the Fig.~\ref{11}(b) shows the variation of $I_C/(\hbar \Omega)^2$ with exchange coupling $J_{\rm T}$ and precession angle $\alpha$. 

Using similar approach as in Refs.~\cite{Vavilov,adiab}, we can obtain a better understanding of the quadratic dependence of charge current on frequency. In our single-parameter pumping setup, the in-plane $x,y$ components of the magnetization rotate periodically, tracing out an elliptic contour in the two-dimensional phase space of the pumping coordinate $X$ and its time derivative $ \dot X$. The charge pumped per cycle is equal to the area enclosed by this contour, which for an ellipse grows linearly with the driving frequency \(\Omega\). Simultaneously, the number of cycles per unit time is proportional to \(\Omega\). Consequently, the resulting charge current scales quadratically with the driving frequency.

An analytic result for the charge current in the adiabatic regime can be similarly obtained as for the spin current in Eq.~\eqref{IsEX}, yielding 
\begin{eqnarray}\label{IcEX}
&&I_{C}=\frac{4\Gamma e|V|^2m\Omega^2 \sin^2{\alpha}}{\hbar^3J_{\rm T}(J_{\rm T}^2+4\Gamma^2)^2(\Gamma^2+\mu^2)}\bigg\{-4J_{\rm T}^3\Gamma\nonumber\\
&&-16J_{\rm T}\Gamma^3-\pi(J_{\rm T}^2-4\Gamma^2)(\Gamma^2+\mu^2)\\
&&+2(\Gamma^2+\mu^2)\bigg[(J_{\rm T}^2-4\Gamma^2)\cot^{-1}{\frac{2J_{\rm T}\Gamma}{\mu^2+\Gamma^2-J_{\rm T}^2}}\nonumber\\
&&+2J_{\rm T}\Gamma\log{\frac{J_{\rm T}^4+2J_{\rm T}^2(\Gamma^2-\mu^2)+(\Gamma^2+\mu^2)^2}{(\Gamma^2+\mu^2)^2}}\bigg]\bigg\},\nonumber
\end{eqnarray}
which shows that up to leading-order $I_C$ is indeed quadratic in $\Omega$. Both analytic results  Eqs.~\eqref{IsEX}-\eqref{IcEX} also confirm the $\sin^2{\alpha}$ dependence on the precession angle $\alpha$ as originally concluded from the numerical results in Fig.~\ref{3} and Fig.~\ref{9}. 

The effect of spin-orbit coupling for our system is found to be small in the low-frequency regime, as the lowest corrections only appear in the second order in the spin-orbit coupling strength. Eqs.~\eqref{IsEX}-\eqref{IcEX} therefore also apply to the case of spin-degenerate 2DEGs with a parabolic energy dispersion. In addition, we notice that both currents Eqs.~\eqref{IsEX}-\eqref{IcEX} are proportional to the broadening parameter $\Gamma$, and thus vanish in the limit $\Gamma \to 0$. This implies that a finite quasiparticle lifetime is essential for the tunneling currents in the adiabatic regime.

\begin{figure}
   \begin{center}
            \includegraphics[width=\columnwidth]{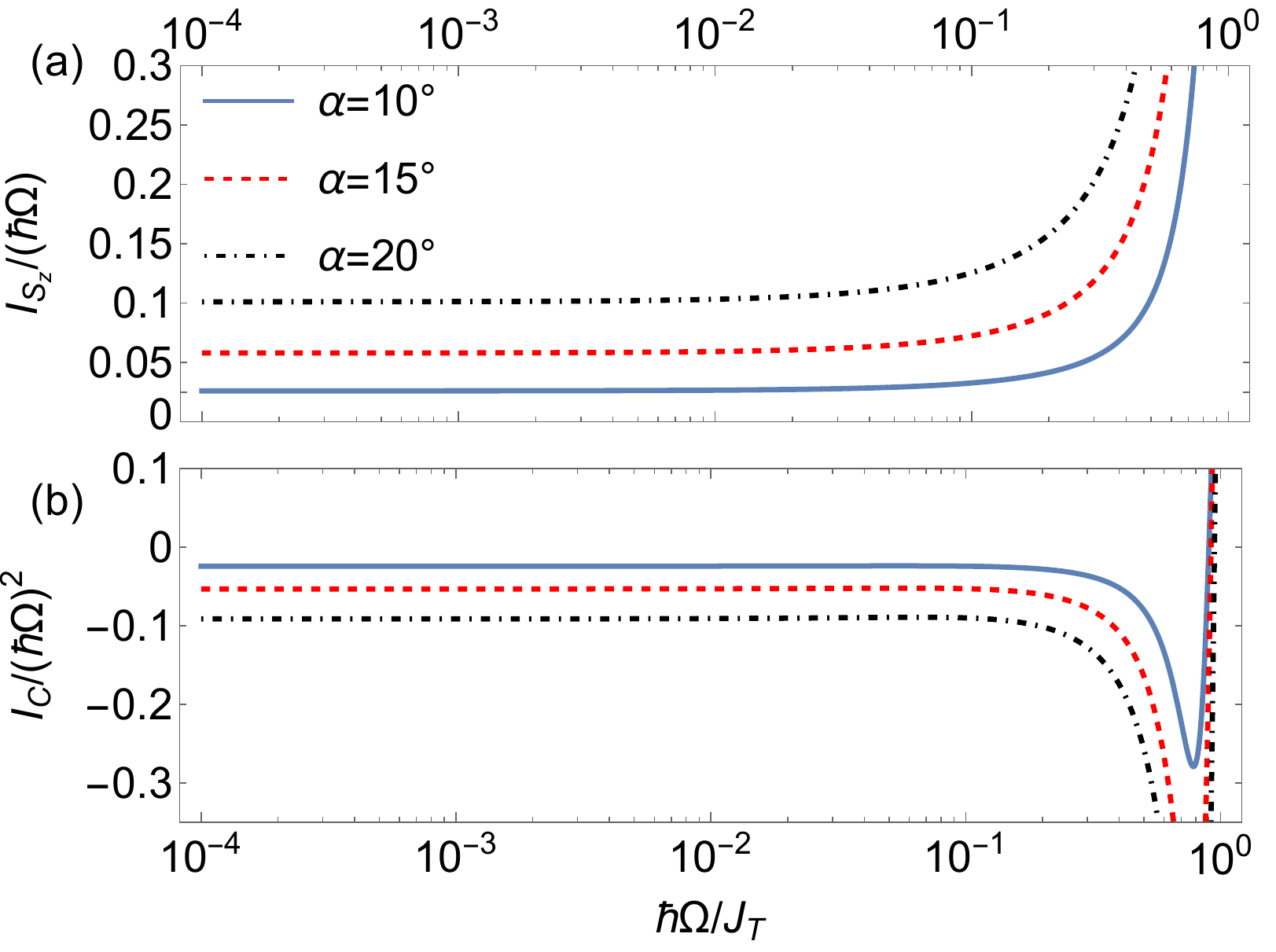}
                \end{center}
                \caption{Linear-log plot of time-averaged tunneling current $I_{S_z}/\hbar \Omega$ and $|I_C|/(\hbar \Omega)^2$ as a function of  $\hbar \Omega/J_{\rm T}$ at precession angles $\alpha=10\degree, 15\degree, 20\degree$. The exchange coupling is $J_{\rm T}=10\,{\rm  meV}$, $\Gamma=1{\rm mev}$ and the other parameters $, V,\mu$ are the same as in Fig.~\ref{3}.}\label{ILinLog}

\end{figure}
\subsection{Non-adiabatic Regime} \label{nonadia}

Our exact Floquet formulation allows us to go beyond the adiabatic regime to situations where $\hbar\Omega \sim J_{\rm T}$, which can happen when the driving frequency is high enough or when the exchange coupling is small enough. At this point, the adiabatic assumption falls apart and the conventional scattering approach \cite{revpump} to spin pumping becomes inadequate. To examine how the conventional adiabatic pumping behavior breaks down, 
we calculate the time-averaged spin and charge currents using Eqs.~\eqref{Is}-\eqref{Ic} for an extended range of frequency at a fixed value of  $J_{\rm T}$. The linear-log plot in Fig.~\ref{ILinLog}(a) show the scaled spin current $I_{S_z}/(\hbar \Omega)$ as a function of $\hbar\Omega/J_{\rm T}$ up to $1$ for $J_{\rm T} = 10\,\mathrm{meV}$ and for different values of $\alpha$. The adiabatic regime is indicated by the flat regions at smaller values of  $\hbar\Omega/J_{\rm T}$. As the frequency is increased above $\hbar\Omega/J_{\rm T} \sim 0.1$, one can clearly see that the plots start to deviate noticeably from the flat regions, corresponding to $I_{S_z}$ showing a nonlinear behavior versus $\hbar\Omega$. 
Similar observations can be made for the charge current displayed in Fig.~\ref{ILinLog}(b), where the deviations from the flat regions correspond to a non-quadratic behavior of $I_C$ versus $\hbar\Omega$. 
These deviations at higher driving frequencies from the linear and quadratic dependence signal the breakdown of the adiabatic regime where the conventional pumping theory fails, but are fully captured within our Floquet-based theory. 

The sharp deviation for both $I_{S_z}/(\hbar \Omega)$ and $I_{C}/(\hbar \Omega)^2$ near $\hbar\Omega/J_{\rm T} = 1$ requires a closer examination. Fig.~\ref{ILinLin}(a)-(b) show the spin current and the charge current on a linear scale up to $\hbar\Omega/J_{\rm T} = 2$. Now one can see that the increase is actually part of a peak near $\hbar\Omega/J_{\rm T} = 1$, with the peak position changing with different values of $\alpha$. We can understand the origin of this peak if we look at the expression of the spin and charge currents in Eqs.~\eqref{Is}-\eqref{Ic}. This peak originates from the term in the second contribution $\sim \sin^2\theta$ containing a product of the spectral functions. 
This is best illustrated by considering the limit of zero broadening $\Gamma \to 0$, 
upon which the last term $\propto \Gamma$ in each of Eqs.~\eqref{Is}-\eqref{Ic} vanishes. Then the spectral function becomes a Dirac delta function, and the $\omega$-integration results in another Dirac delta function that requires $E^{\rm T}_{q,\pm} = E^{\rm B}_{q,\pm}$. This implies a resonant tunneling condition between the magnon-dressed quasiparticles in the top layer and the bottom layer. If we further consider vanishing spin-orbit coupling $\lambda = 0$, this condition further simplifies becoming independent of $q$, yielding a delta-function peak located at 
%
%
\begin{equation} \label{peak}
\frac{\hbar\Omega}{J_{\rm T}} =  \frac{1}{\cos\alpha}.     
\end{equation}
This simple formula provides a good approximation to the  location of the peaks observed in our numerical results since $\lambda/\Delta \ll 1$, as seen in Table~\ref{peakdata}. In the limit $\Gamma = \lambda = 0$ we are considering, the peak locations for the spin current and the charge current coincide, with both given by Eq.~\ref{peak}. When $\Gamma$ and $\lambda$ become nonzero, their peak positions become slightly different as observed in Fig.~\ref{ILinLin}, but are still both reasonably well approximated by Eq.~\ref{peak}. 

\begin{table}[ht] \label{peakdata}
\centering 
\begin{tabular}{c c c c} 
\hline\hline 
 $\alpha$ & Numerical $I_{S_z}$ & Numerical $I_C$ & Analytical Eq.~\ref{peak} \\ [0.5ex] 
\hline 
$5\degree$ & $1.025$ & $1.00$ & $1.00$\\
$10\degree$ & $1.04$ & $1.01$ & $1.02$ \\ 
$15\degree$ & $1.06$ & $1.03$ & $1.04$ \\
$20\degree$ & $1.08$ & $1.05$ & $1.06$ \\ [1ex] 
\hline 
\end{tabular}
\label{table:nonlin} 
\caption{Values of the peak position in the non-adiabatic regime $\hbar\Omega \sim J_{\rm T}$ for precession angles $\alpha = 5\degree, 10\degree, 15\degree, 20\degree$. The first two columns show the values obtained from numerical calculations of $I_{S_z}/(\hbar\Omega)$ and $I_C/(\hbar\Omega)^2$, while the last column shows the analytically obtained values from Eq.~\eqref{peak}.} 
\end{table}
To better understand the origin and location of the above discussed peak, using the definition of the spectral functions in Eq.~(\ref{spectralfcn})  we can rewrite Eqs.~(\ref{Is})-(\ref{Ic}) in the following form as a product of three Lorentzians. Here we assumed that the top and bottom layers have same chemical potential.
\begin{eqnarray}\label{IsA}
&&I_{S_z}=\frac{2e\Gamma}{\hbar\pi}|V|^2 \sum_{q\tau\tau'}\sum_{\gamma\beta}J_T^2\sin^2\alpha\int_{-\infty}^{\infty} d\omega\,\gamma\nonumber\\
&&\times\left[f_{\rm T}(\omega+\gamma\frac{\Omega}{2})-f_{\rm T}(\omega-\gamma\frac{\Omega}{2})\right]\frac{1}{(\omega-E^{\rm B}_{q,\gamma})^2+\Gamma^2}\nonumber\\
&&\times\frac{1}{(\omega-E^{\rm T}_{q,+})^2+\Gamma^2}\frac{1}{(\omega-E^{\rm T}_{q,-})^2+\Gamma^2},
\end{eqnarray}
\begin{eqnarray}\label{IcA} 
&&I_{C}=\frac{2e\Gamma}{\hbar\pi}|V|^2 \sum_{q\tau\tau'}\sum_{\gamma\beta}J_T^2\sin^2\alpha\int_{-\infty}^{\infty} d\omega\,\nonumber\\
&&\times\left[f_{\rm T}(\omega+\gamma\frac{\Omega}{2})-f_{\rm T}(\omega-\gamma\frac{\Omega}{2})\right]\frac{1}{(\omega-E^{\rm B}_{q,\gamma})^2+\Gamma^2}\nonumber\\
&&\times\frac{1}{(\omega-E^{\rm T}_{q,+})^2+\Gamma^2}\frac{1}{(\omega-E^{\rm T}_{q,-})^2+\Gamma^2}.
\end{eqnarray}

From the above equations, we can see that the three spectral functions in Eqs.~(\ref{IsA})-(\ref{IcA}) are in the form of Lorentzians centering around $\omega=E^{\rm T}_{q,\pm}$ and $\omega=E^{\rm B}_{q,\gamma}$, where $\gamma = \pm$. The distance between the positions of these Lorentzians, which controls their mutual  overlap, together with the overall factor of $J_T^2\sin^2\alpha$ determine the magnitude of the spin and charge currents. From the expressions of the energies in Eq.~(\ref{HTnu2}) and Eq.~(\ref{HBnu2}), we can infer that for small $\lambda/\Delta$ the energies $E^{\rm T}_{q,\pm}$ become approximately the same as $E^{\rm B}_{q,\gamma}$, making the top layer Lorentzians overlap with the bottom layer Lorentzian. The peak of the spin and charge currents at $J_T=\hbar\Omega\cos\alpha$ occurs when one Lorentzian of top layer overlaps with the Lorentzian of bottom layer and their centers coincide, maximizing the integral. 

The previously described frequency‐dependent peak is strongly suppressed in the adiabatic regime (Fig.~\ref{IS_J}-\ref{IC_J}) because the frequency (on the order of $\mu{\rm eV}$) is much smaller than the broadening $\Gamma$ ($\sim{\rm meV}$). Instead, a distinct peak appears as $J_T$ is varied (see Fig.~\ref{IS_J}-\ref{IC_J}).  To understand this peak, note from the expression for $E^{\rm T}_{q,\pm}$ that increasing $J_T$ enlarges the separation between the three Lorentzians. In the adiabatic regime, focusing only on the linear-in-$\Omega$ term and omitting the small $\lambda/\Delta$ correction, the pumped current can be approximated as an integral over three Lorentzians centering at $\mu$ and $\mu\pm J_T$. The result of the integral scales as $J_T^2/[(J_T^2+\Gamma^2)(J_T^2+4\Gamma^2)]$. This is consistent with what can be derived from our approximate analytic result in Eq.~(\ref{IsEX}) when the $J_T^2$ terms dominate in the square bracket.

When $J_T$ is much smaller than $\Gamma$ that  gives the broadening of the Lorentzians, the Lorentzian peaks strongly overlap with each other. In such a limit, the $J_T^2$ factor in the numerator dominates and the magnitude of the currents grows like $J_T^2$. When $J_T\gg \Gamma$, $J_T$ becomes the primary term in the denominator of $\tilde{\mathcal A}_{\rm T,\tau\pm}$ and the currents decay like $1/J_T^2$. Therefore, one expects a maximum to occur in the intermediate range when $J_T\sim\Gamma$. This can be confirmed by analytically finding the location of the maximum for the expression $J_T^2/[(J_T^2+\Gamma^2)(J_T^2+4\Gamma^2)]$, given by the integrated result of the three approximated Lorentzians centering at $\mu$ and $\mu\pm J_T$ as explained before. This gives a maximum at $J_T=\sqrt{2} \Gamma$, which is consistent with our observation in Fig.~\ref{IS_J}-\ref{IC_J}.

\begin{figure}
   \begin{center}
            \includegraphics[width=\columnwidth]{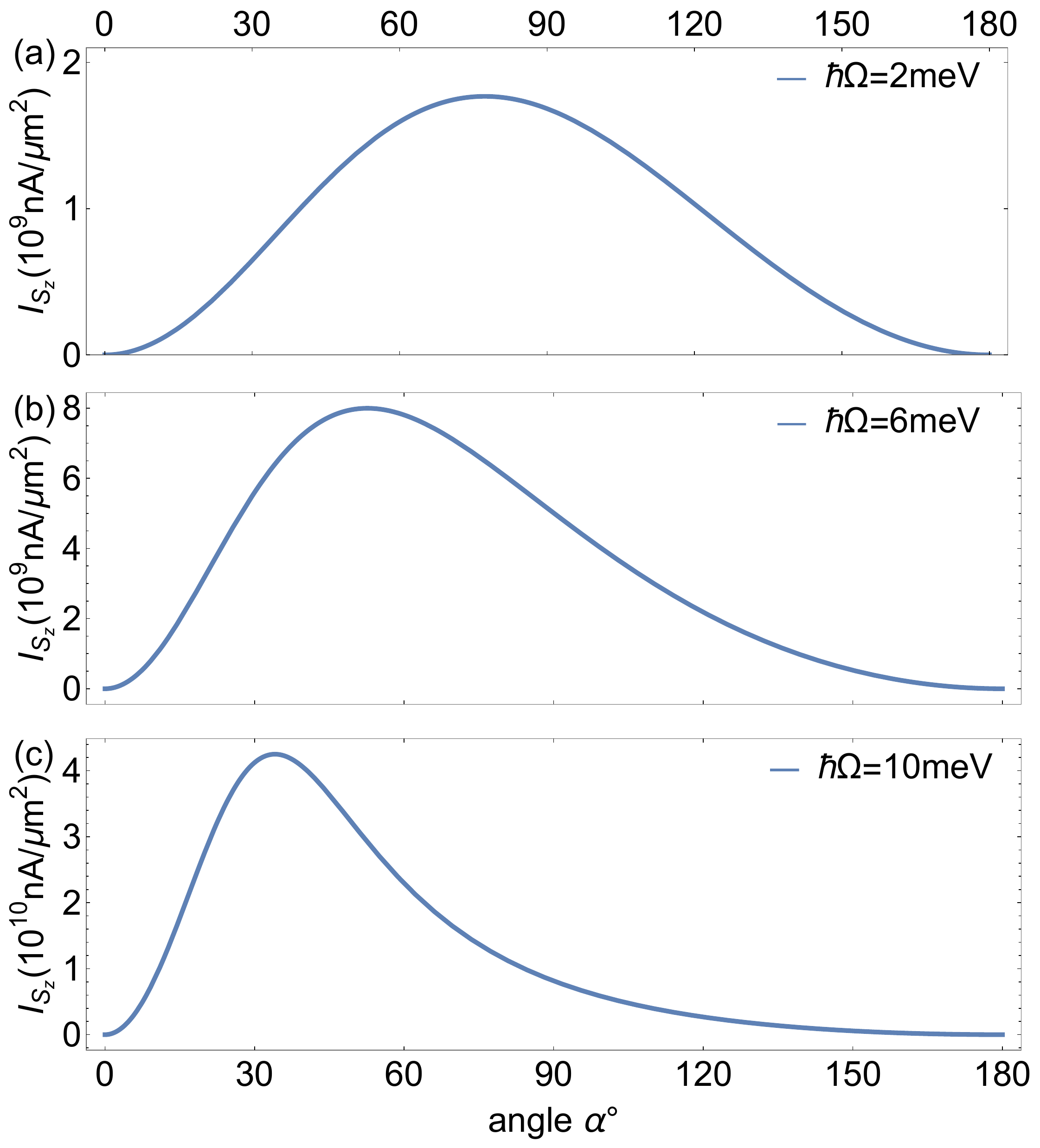}
                \end{center}
                \caption{Time-averaged tunneling spin current $I_{S_z}$ as a function of precession angle $\alpha$ with  at driving frequencies $\hbar\Omega=2,6,10\,{\rm m eV}$.  The exchange coupling is $J_{\rm T}=10\,{\rm  meV}$, and the other parameters  $\Gamma, V,\mu$ are the same as in Fig.~\ref{3}.}\label{ISalpha_NA}

\end{figure}

\begin{figure}
   \begin{center}
            \includegraphics[width=\columnwidth]{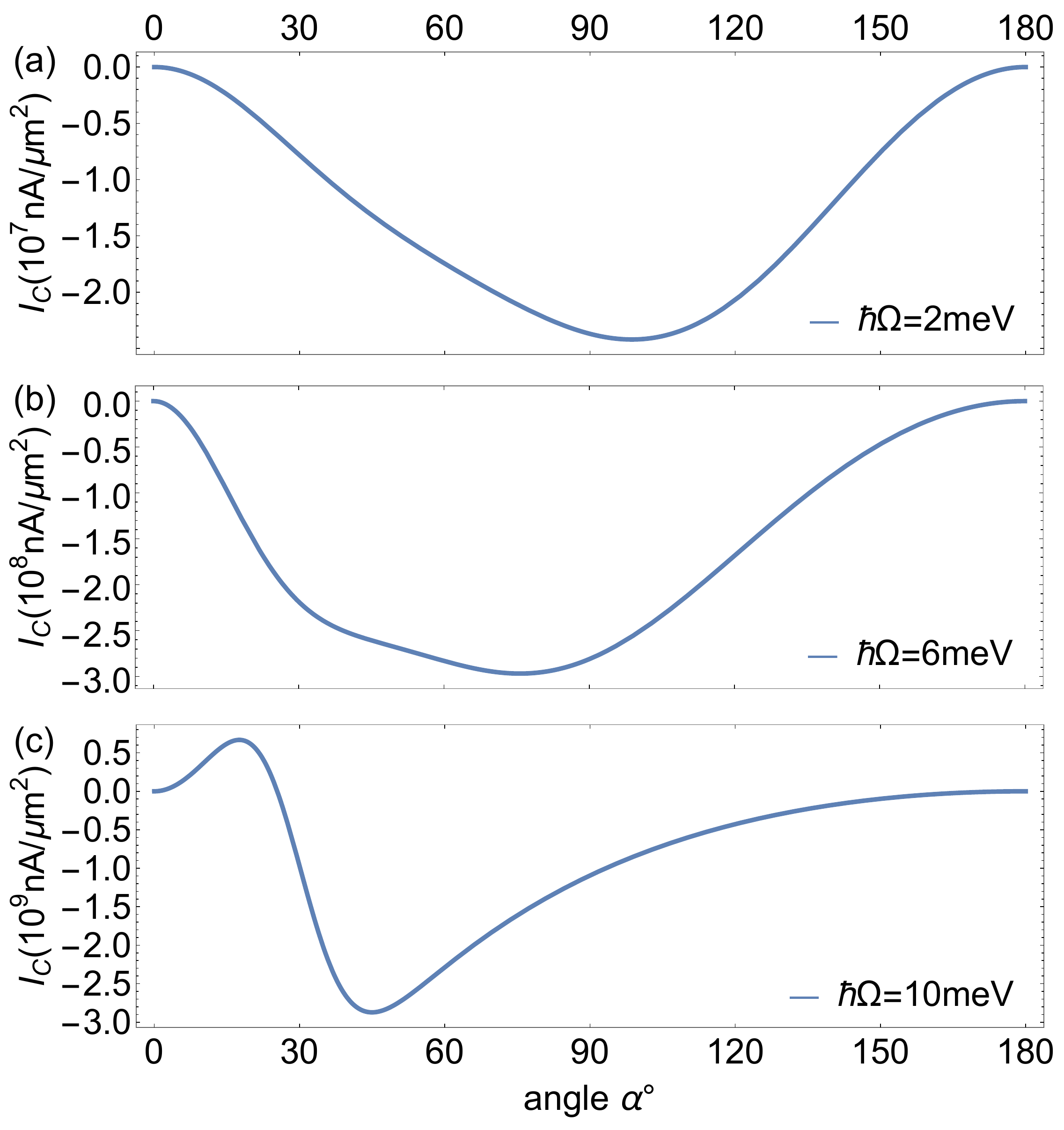}
                \end{center}
                \caption{Time-averaged tunneling spin current $I_C$ as a function of precession angle $\alpha$  at driving frequencies  $\hbar\Omega=2,6,10\, {\rm m eV}$. The exchange coupling is $J_{\rm T}=10\,{\rm  meV}$, and the other parameters  $\Gamma, V,\mu$ are the same as in Fig.~\ref{3}.}\label{ICalpha_NA}

\end{figure}

We finally look at the dependence of the spin and charge currents on the precession angle. Fig.~\ref{ISalpha_NA} shows the spin current $I_{S_z}$ as a function of $\alpha$ for different driving frequencies $\hbar\Omega = 2, 6, 10\,\mathrm{meV}$ and a fixed exchange coupling $J_{\rm T} = 10\,\mathrm{meV}$. As $\hbar\Omega/J_{\rm T}$ increases from panel (a) to (c), it is seen that the $\alpha$-dependence becomes different from $\sin^2\alpha$ that was found in the adiabatic regime, with the peak position shifting towards smaller angles. For the  charge current $I_C$ displayed in Fig.~\ref{ICalpha_NA}, the difference from $\sin^2\alpha$ dependence is even more noticeable, and at $\hbar \Omega = 10\,\mathrm{meV}$ in particular, $I_C$ exhibits a sign change turning positive at smaller angles. 

An important implication of our results is that in the non-adiabatic regime, where the driving frequency is comparable to or exceeds the interfacial exchange coupling $ \hbar\Omega \gtrsim J_T$, the pumped spin current can surpass the critical value required for magnetization switching in a typical thin-film ferromagnet. In such systems with an  injected spin current, the threshold spin current density for switching is determined by the balance between the spin-transfer torque exerted by the injected spins and the damping torque of the ferromagnet.  Using typical parameters for a cobalt (Co) layer---one of the most commonly used materials in the literature for spin-torque switching---together with literature values of the saturation magnetization \(M_s\), Gilbert damping parameter \(\alpha\) and thickness~\cite{Katine1,Katine2}, the critical spin current density per unit area is estimated to be on the order of \(10^{8} \,\mathrm{nA}/\mu\mathrm{m}^2\). Our calculations show that within the adiabatic regime, the spin current remains below this threshold. However, once the driving frequency exceeds the meV scale, as in the high-frequency range of Fig.~\ref{ISalpha_NA}, the spin current can reach \(\sim 10^{9} \,\mathrm{nA}/\mu\mathrm{m}^2\), clearly exceeding the switching threshold. This demonstrates that non-adiabatic spin pumping enabled by high-frequency driving can generate spin currents strong enough to induce magnetization reversal in thin ferromagnetic layers.

%
%

The various features predicted in this subsection for the non-adiabatic regime could in principle be realized using an antiferromagnet in the setup as discussed in Sec.~\ref{Formulation}. Because typical AFMR frequencies can reach up to $\sim 10 \,\mathrm{meV}$ that is covered by 
the range of the interfacial exchange coupling $J_{\rm T} \sim 10\,\mathrm{meV}\,-\,1\,\mathrm{eV}$, for small $J_{\rm T} \sim 10\,\mathrm{meV}$ the non-adiabatic regime $\hbar\Omega/J_{\rm T} \sim 1$ should be realizable. Since the resonance frequency $\Omega$ and the precession angle $\alpha$ \cite{cone1,cone2,cone3} can be obtained experimentally, measurement of the peak location of the tunneling spin current or the charge current could also offer a new method to estimate 
the interfacial exchange coupling $J_{\rm T}$ using the approximate analytic formula  Eq.~\eqref{peak}.

\section{Conclusion}\label{conclusion}
In this paper, we have investigated the tunneling of spin current and the accompanying charge current due to spin pumping through a tunneling heterostructure that is coupled to a magnetic layer driven at resonance. By employing the Floquet-Keldysh Green’s function formalism, we developed a non-perturbative approach to calculate the tunneling spin current and charge current driven by the magnetization precession of the magnetic layer. Our analysis reveals how the tunneling spin current is influenced by system parameters such as the precession angle, driving frequency, and interfacial exchange coupling. The theoretical framework presented in this paper is applicable across both the adiabatic pumping regime at microwave frequencies and higher frequency regimes beyond the microwave range. In the low-frequency regime, we derived analytical expressions for the tunneling spin and charge currents, whose dependence on the driving frequency and the precession angle agrees with the behaviors in the adiabatic regime. 
Our full numerical analysis demonstrates significant deviations from these conventional behaviors when the ratio of driving frequency to exchange coupling is greater than about $0.1$. In particular, we find that resonance tunneling can occur which is signified by the occurrence of a peak in the spin current and the charge current.  
While ferromagnetic resonance generally operates within the microwave frequency range, our setup has the potential to reach terahertz frequencies through antiferromagnetic resonance. This capability allows for the exploration of spin pumping phenomena at higher frequencies, opening new avenues for research and applications in antiferromagnetic spintronics. 
\\

\acknowledgments
We thank Y. Araki, J.~B. Mohammadi, and F. Xue  for useful  discussions. This work was supported by the U.S. Department of Energy, Office of Science, Basic Energy Sciences under Early Career Award No. DE-SC0019326 (M.K and W-K.T.) and by the National Science Foundation via Grant No. DMR-2213429 (M.M.A.). 
\section*{Data Availability}

Results were visualized in Mathematica 13.0~\cite{mathematica13}~, the data that support the findings of this article are openly available~\cite{data}.
\appendix{}

\section{Derivation of the Single-Layer-Driven Limit from the Double-Layer-Driven Case}
In this appendix we present the derivation of reducing Eq.~(\ref{current4})  to Eq.~(\ref{Ic}) when only one layer is driven. When the bottom layer is not driven, we take the limit $\theta_{\rm B}\rightarrow0$ and $J_{\rm B}\rightarrow0$ and Eq.~(\ref{current4}) become

\begin{eqnarray}\label{currentapp1}
&&I_{\rm C}=-\frac{e\pi}{2\hbar}|V|^2 \sum_{q\tau\tau'}\sum_{\gamma\beta S}\int_{-\infty}^{\infty} d\omega \bigg\{\tilde{\mathcal A}_{\rm T,\tau\beta}(\omega)\tilde{\mathcal A}_{\rm B,\tau'\gamma}(\omega)\nonumber\\
&&\times\bigg[(1+\gamma\beta\cos{\theta_{\rm T}})[f_{\rm T}(\omega+S\frac{\Omega}{2})(1+\beta S\cos{\theta_{\rm T}})\\
&&-f_{\rm B}(\omega+S\frac{\Omega}{2})(1+\gamma S)]+\gamma S\sin^2{\theta_{\rm T}}f_{\rm T}(\omega+S\frac{\Omega}{2})]\bigg]\nonumber\\
&&-2\frac{\pi}{\Gamma}\sin^2{\alpha}\ \gamma S\tilde{\mathcal A}_{\rm T,\tau+}(\omega)\tilde{\mathcal A}_{\rm T,\tau-}(\omega)\tilde{\mathcal A}_{\rm B,\tau'\gamma}(\omega)J_T^2f_{\rm T}(\omega+S\frac{\Omega}{2})\bigg\},\nonumber
\end{eqnarray}

where the term proportional to $J_B^2/\sin{\theta_{\rm B}}$ in Eq.~(\ref{current4})  vanishes as $J_B^2/\sin{\theta_{\rm B}}\propto J_{\rm B}$ which goes to zero in our limit.

The above equation can then be rewritten into the following form
\begin{eqnarray}\label{currentapp2}
&&I_{\rm C}=-\frac{e\pi}{2\hbar}|V|^2 \sum_{q\tau\tau'}\sum_{\gamma\beta S}\int_{-\infty}^{\infty} d\omega \bigg\{\tilde{\mathcal A}_{\rm T,\tau\beta}(\omega)\tilde{\mathcal A}_{\rm B,\tau'\gamma}(\omega)\nonumber\\
&&\times\left[f_{\rm T}(\omega+S\frac{\Omega}{2})[1+(\beta S+\gamma\beta)\cos{\theta_{\rm T}}+\gamma S]-f_{\rm B}(\omega+S\frac{\Omega}{2})\right.\nonumber\\
&&\left.(1+\gamma S)(1+\gamma\beta\cos{\theta_{\rm T}})+\gamma S\sin^2{\theta_{\rm T}}f_{\rm T}(\omega+S\frac{\Omega}{2})]\right]\\
&&-2\frac{\pi}{\Gamma}J_T^2\sin^2{\alpha}\ \gamma S\tilde{\mathcal A}_{\rm T,\tau+}(\omega)\tilde{\mathcal A}_{\rm T,\tau-}(\omega)\tilde{\mathcal A}_{\rm B,\tau'\gamma}(\omega)f_{\rm T}(\omega+S\frac{\Omega}{2})\bigg\}.\nonumber
\end{eqnarray}

We can then split the terms into $S=\gamma$ and $S=-\gamma$ and take the sum over $S$ to get

\begin{eqnarray}
&&I_{\rm C}=-\frac{e\pi}{\hbar}|V|^2 \sum_{q\tau\tau'}\sum_{\gamma\beta}\int_{-\infty}^{\infty} d\omega \bigg\{(1+\gamma\beta\cos{\theta})\nonumber\\
&&\times\tilde{\mathcal A}_{\rm T,\tau\beta}(\omega)\tilde{\mathcal A}_{\rm B,\tau'\gamma}(\omega)\left[f_{\rm T}(\omega+\gamma\frac{\Omega}{2})-f_{\rm B}(\omega+\gamma\frac{\Omega}{2})\right]\nonumber\\
&&-\frac{\pi}{\Gamma}J_T^2 \sin^2{\alpha}\ \tilde{\mathcal A}_{\rm T,\tau+}(\omega)\tilde{\mathcal A}_{\rm T,\tau+}(\omega)\tilde{\mathcal A}_{\rm B, \tau'\gamma}(\omega)\nonumber\\
&&\times\left[f_{\rm T}(\omega+\gamma\frac{\Omega}{2})-f_{\rm T}(\omega-\gamma\frac{\Omega}{2})\right]\bigg\},
\end{eqnarray}
which is same as Eq.~(\ref{Ic}) in the main text for the single-layer-driven case.

\bibliography{refs}
\end{document}